\documentclass[lettersize,journal]{IEEEtran}
\usepackage{amsmath,amsfonts}
\usepackage{algorithmic}
\usepackage{algorithm}
\usepackage{array}
\usepackage[caption=false,font=normalsize,labelfont=sf,textfont=sf]{subfig}
\usepackage{textcomp}
\usepackage{stfloats}
\usepackage{url}
\usepackage{verbatim}
\usepackage{graphicx}
\usepackage{ulem} 
\usepackage{cite}
\usepackage{float}
\usepackage{booktabs}
\usepackage{multirow}
\usepackage{cite}
\usepackage{stfloats}
\usepackage{amsmath}
\usepackage{bbding}
\usepackage{ulem}
\usepackage{xcolor}

\usepackage{tikz,xcolor} 
\usepackage[implicit=false]{hyperref}

\definecolor{lime}{HTML}{A6CE39}
\DeclareRobustCommand{\orcidicon}{%
	\begin{tikzpicture}
	\draw[lime, fill=lime] (0,0) 
	circle [radius=0.16] 
	node[white] {{\fontfamily{qag}\selectfont \tiny ID}};
	\draw[white, fill=white] (-0.0625,0.095) 
	circle [radius=0.007];
	\end{tikzpicture}
	\hspace{-2mm}
}

\foreach \x in {A, ..., Z}{%
	\expandafter\xdef\csname orcid\x\endcsname{\noexpand\href{https://orcid.org/\csname orcidauthor\x\endcsname}{\noexpand\orcidicon}}
}

\usepackage{xspace}
\makeatletter
\DeclareRobustCommand\onedot{\futurelet\@let@token\@onedot}
\def\@onedot{\ifx\@let@token.\else.\null\fi\xspace}
\def\eg{\textit{e.g\onedot}} 
\def\ie{\textit{i.e\onedot}} 

\makeatother

\usepackage{lineno}

\begin{document}

\title{Annular Computational Imaging: Capture Clear Panoramic Images through Simple Lens}

\author{Qi Jiang†, Hao Shi†, Lei Sun, Shaohua Gao, Kailun Yang, and Kaiwei Wang
\thanks{This work was supported by the National Natural Science Foundation of China (NSFC) under Grant No. 12174341}
\thanks{Qi Jiang, Hao Shi, Lei Sun, Shaohua Gao, and Kaiwei Wang are with State Key Laboratory of Modern Optical Instrumentation, Zhejiang University, China \{qijiang, haoshi, leo\_sun, gaoshaohua, wangkaiwei\}@zju.edu.cn}
\thanks{Kailun Yang is with Institute for Anthropomatics and Robotics, Karlsruhe Institute of Technology, Germany {kailun.yang@kit.edu}}
\thanks{Lei Sun is also with ETH Z\"urich, Switzerland {leisun@ee.ethz.ch}}%
\thanks{Corresponding author: Kaiwei Wang.}%
\thanks{†These authors contributed equally.}%
}

\markboth{IEEE TRANSACTIONS ON COMPUTATIONAL IMAGING, June~2022}%
{Shell \MakeLowercase{\textit{et al.}}: A Sample Article Using IEEEtran.cls for IEEE Journals}

\IEEEpubid{}

\maketitle

\begin{abstract}

Panoramic Annular Lens (PAL) composed of few lenses has great potential in panoramic surrounding sensing tasks for mobile and wearable devices because of its tiny size and large Field of View (FoV). However, the image quality of tiny-volume PAL confines to optical limit due to the lack of lenses for aberration correction. In this paper, we propose an Annular Computational Imaging (ACI) framework to break the optical limit of light-weight PAL design. To facilitate learning-based image restoration, we introduce a wave-based simulation pipeline for panoramic imaging and tackle the synthetic-to-real gap through multiple data distributions. The proposed pipeline can be easily adapted to any PAL with design parameters and is suitable for loose-tolerance designs. Furthermore, we design the Physics Informed Image Restoration Network ({PI$^{2}$RNet}) considering the physical priors of panoramic imaging and single-pass physics-informed engine. At the dataset level, we create the DIVPano dataset and the extensive experiments on it illustrate that our proposed network sets the new state of the art in the panoramic image restoration under spatially-variant degradation. In addition, the evaluation of the proposed ACI on a simple PAL with only 3 spherical lenses reveals the delicate balance between high-quality panoramic imaging and compact design. To the best of our knowledge, we are the first to explore Computational Imaging (CI) in PAL. Code and datasets are publicly available at \url{https://github.com/zju-jiangqi/ACI-PI2RNet}.
\end{abstract}


\begin{figure*}[!t]
\centering
\includegraphics[width=1\linewidth]{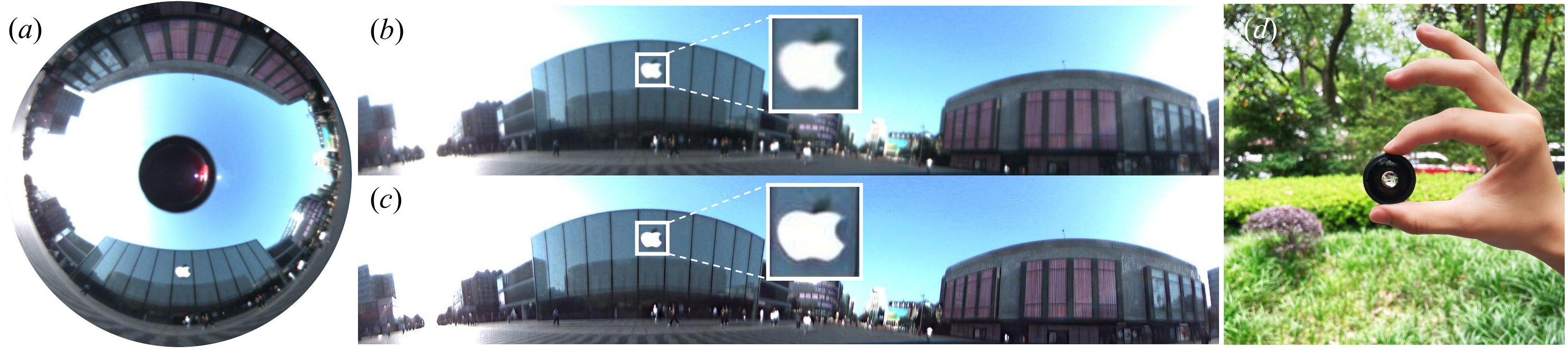}
\vskip-2ex
\caption{We capture a clear panoramic image through simple lens. (a) is the degraded annular image captured by a PAL with tiny size shown in (d). In (b) and (c), we transform (a) into an unfolded image and restore it with our proposed framework. Our Annular Computational Imaging (ACI) improves the image quality of PAL composed of few lenses.}
\label{fig:introduction}
\vskip-3ex
\end{figure*}

\section{Introduction}
\label{sec:Introduction}

\IEEEPARstart{I}{mages}  with very large Field of View (FoV) are attracting more attention on account of more abundant information they provide about the surrounding environment for some Computer Vision (CV) tasks~\cite{gao2022review}. A considerable amount of research is carried out on semantic segmentation~\cite{yang2019pass,yang2019can,sun2021aerial}, object detection~\cite{2005People,2017Object}, visual SLAM~\cite{lin2018pvo,won2020omnislam,2019PALVO} and other CV tasks~\cite{fang2020panoramic,2021UniFuse,shi2022panoflow} under images with 360$^\circ$ FoV, namely panoramic CV tasks. Mainstream camera models (\eg~pinhole camera, fisheye camera) are incapable of panoramic imaging with a FoV of 360$^\circ$. As a feasible solution, these cameras are equipped with multi-view image mosaic algorithms~\cite{2005People,2017Object,lin2018pvo} to make 360$^\circ$ panoramic imaging possible. However, the problems introduced by this sub-optimal choice (\eg~high latency between devices, inaccurate multi-view alignment and heavy computational burden) are unbearable for its application in mobile and wearable devices or some extreme environments.

Panoramic Annular Lens (PAL) is a compact solution to panoramic imaging with 360$^\circ$ FoV but a late starter~\cite{greguss1986panoramic,powell1994panoramic}. Due to the catadioptric structure of the PAL head, a PAL can achieve 360$^\circ$ panoramic imaging, where incident rays of 360$^\circ$ FoV are collected and bent to a beam with small angle. To correct aberrations in different FoVs, complex relay lens groups are applied so that traditional PAL systems normally consist of more than 8 lenses \cite{Niu2007ImprovementID,Cheng2016DesignOA,Zhang2020DesignOA}. Simplifying the number of lenses is becoming a hot research topic. Some researchers leverage Q-type surfaces in high-quality panoramic systems design\cite{gao2021design}. However, the special surface design for reduction in the number of lenses brings high cost and difficulties in manufacture, not to mention the still numerous lenses after reduction. Designing PAL with fewer spherical lenses is theoretically possible. But without the lenses 
for aberration correction, the image will degrade seriously, which limits its application. A valuable question emerges: Can we capture clear panoramic images through simple lens?

In the last decade, the great success of convolutional neural network (CNN) promotes the advance of Computational Imaging (CI) methods, in which a main group of efforts attempt to restore the aberration images from simple, imperfect optical lens~\cite{hu2021image,peng2016diffractive,Chenshiqi}, or compact imaging systems with large FoV~\cite{peng2019learned, xue2022deep, yanny2022deep}. Apart from network design, the key to applying CNN is to prepare datasets for network training, which is often realized by live shooting~\cite{2020Replacing,peng2019learned} and optical simulation based on raytracing~\cite{Chenshiqi}, random zernike coefficients~\cite{hu2021image}, optical design software~\cite{li2021universal}, or PSFs calibration~\cite{yanny2022deep, xue2022deep}. However, the imaging simulation of PAL, where the 360° panorama is distributed around the center of the image, is seldom explored in previous CI approaches~\cite{2011Non,2015Blind,heide2016encoded, yanny2022deep, xue2022deep}.

In this paper, to maintain both tiny-volume lens system and high-quality 360$^\circ$ panoramic imaging, we pioneer the Annular Computational Imaging (ACI) framework that can be applied to correct the aberration of any compact PAL design. A simulation pipeline is proposed for panoramic imaging with PAL which generates data pairs for network training based on the aberration distribution in unfolded PAL images and the spatially-variant sampling process during unfolding. The trained CNN can correct optical aberration of given PAL and enables simple lens to capture clear panoramic images.

On the other hand, deep learning based aberration correction~\cite{Chenshiqi,hu2021image} often regards the problem as image restoration which discards the physical model of imaging. To integrate physical priors into restoration, we adopt the idea of single-pass physics-informed engine~\cite{Barbastathis:19} and propose the Physics Informed Image Restoration Network ({\rm PI$^{2}$RNet}). An approximate recovery is made first to consider the distribution of point spread functions (PSFs) and convolution process of imaging, namely physics-informed engine. The followed fine restoration is also improved by degradation characteristics from the physics-informed engine, specially designed convolution kernels and global context components. 

Last but not least, although synthetic datasets~\cite{Chenshiqi,li2021universal} are easier to be mass produced and transferred to different lenses compared to manual shooting~\cite{2020Replacing,peng2019learned}, the synthetic-to-real gap is non-negligible considering that the manufactured lenses often deviate from the theoretical results. In our ACI framework, optical simulation is considered as data augmentation, where the random distributions of PSFs contribute to addressing errors introduced by manufacture and assembly, facilitating loose-tolerance requirements of our framework for lens design. The framework is evaluated on real-world images from a PAL composed of three spherical lenses but with aberration. Finally, we successfully capture clear panoramic images through simple lens, as shown in Fig.~\ref{fig:introduction}. 

The main contributions of our work are in the following:
\begin{itemize}
    \item We propose an Annular Computational Imaging (ACI) framework based on the physical model of panoramic imaging, improving the imaging quality of PAL while keeping its compact structure.
    \item The Physics Informed Image Restoration Network ({\rm PI$^{2}$RNet}) is designed to restore the degraded image, considering the physics-informed engine and spatial distribution of PSFs.
   \item We create the DIVPano dataset through wave-based simulation pipeline, in which random distributions of PSFs are adopted to bridge the synthetic-to-real gap.
    \item Finally, we test the qualitative effects of ACI on a real PAL composed of only three spherical lenses.
\end{itemize}

\section{Related Work}
\label{sec:Related Works}

\subsection{Aberration Correction through Computation}
Simple optical systems, which are inherently flawed because of the compact design, produces sub-optimal images with aberrations. In these cases, computational aberration correction is of great significance, and among these methods, the deconvolution approaches such as the Richardson-Lucy algorithm~\cite{1972Bayesian,lucy1974iterative} and Wiener filter~\cite{1964Extrapolation} are classical approaches. Since the degradation caused by optical aberration is usually spatially-variant, \ie~the linear shift variant (LSV), the traditional hypothesis that blur kernel is invariant is no longer applicable. Trussell and Hunt~\cite{trussell1978image,1978Sectioned} first propose the idea of patch-wise restoration to address the problem. More work~\cite{kim2002curve,bar2007restoration} is then done to alleviate fringe and ringing effects of patch-wise restoration. 

Another option is to calibrate the non-uniform PSFs of the LSV system. Early works~\cite{2008PSF,2011Modeling,2011Non} usually revolve around calibration process based on parameters of given lens and physical aberration model. Additionally, the idea of low-rank decomposition is applied to describe the PSFs of LSV system, which approximates the spatially varying PSFs as weighted sum of shift-invariant kernel~\cite{yanny2022deep, xue2022deep,maalouf2011fluorescence, denis2015fast}. The calibrated PSFs are then used in image restoration. 

As data-driven deep learning methods are extensively applied in image restoration~\cite{7298677,tao2018scale,zhang2019deep}, more aberration correction works are equipped with powerful CNN. These approaches often utilize networks with encoder-decoder structure~\cite{metzler2020deep} combined with efficient blocks for complex spatial distribution of aberrations~\cite{2020Replacing,hu2021image,Chenshiqi,peng2019learned,sun2021end}, obtaining more efficient and robust results than traditional methods. Other networks for image restoration with spatially diverse degradation are also fit for aberration correction which fuse multi-scale information~\cite{tao2018scale,xu2021edpn}, and use deformable convolution~\cite{dai2017deformable,zhu2019deformable} or dynamic convolution~\cite{zhou2019spatio}. Nevertheless, these methods are usually evaluated only on datasets and rarely applied to image restoration of real lenses.

Although computational imaging applied to perspective images captured by pinhole cameras is well developed, the solution to PAL with convenient design is hardly explored. At the same time, physical information corresponding to the imaging model is proved to be a vital factor in aberration-induced degradation~\cite{yanny2022deep, xue2022deep}. In our ACI framework, to solve the problems of image recovery under complex degradation, the network is designed based on the specific distribution of degradation in unfolded PAL images.

\begin{figure*}[!t]
\centering
\includegraphics[width=1\linewidth]{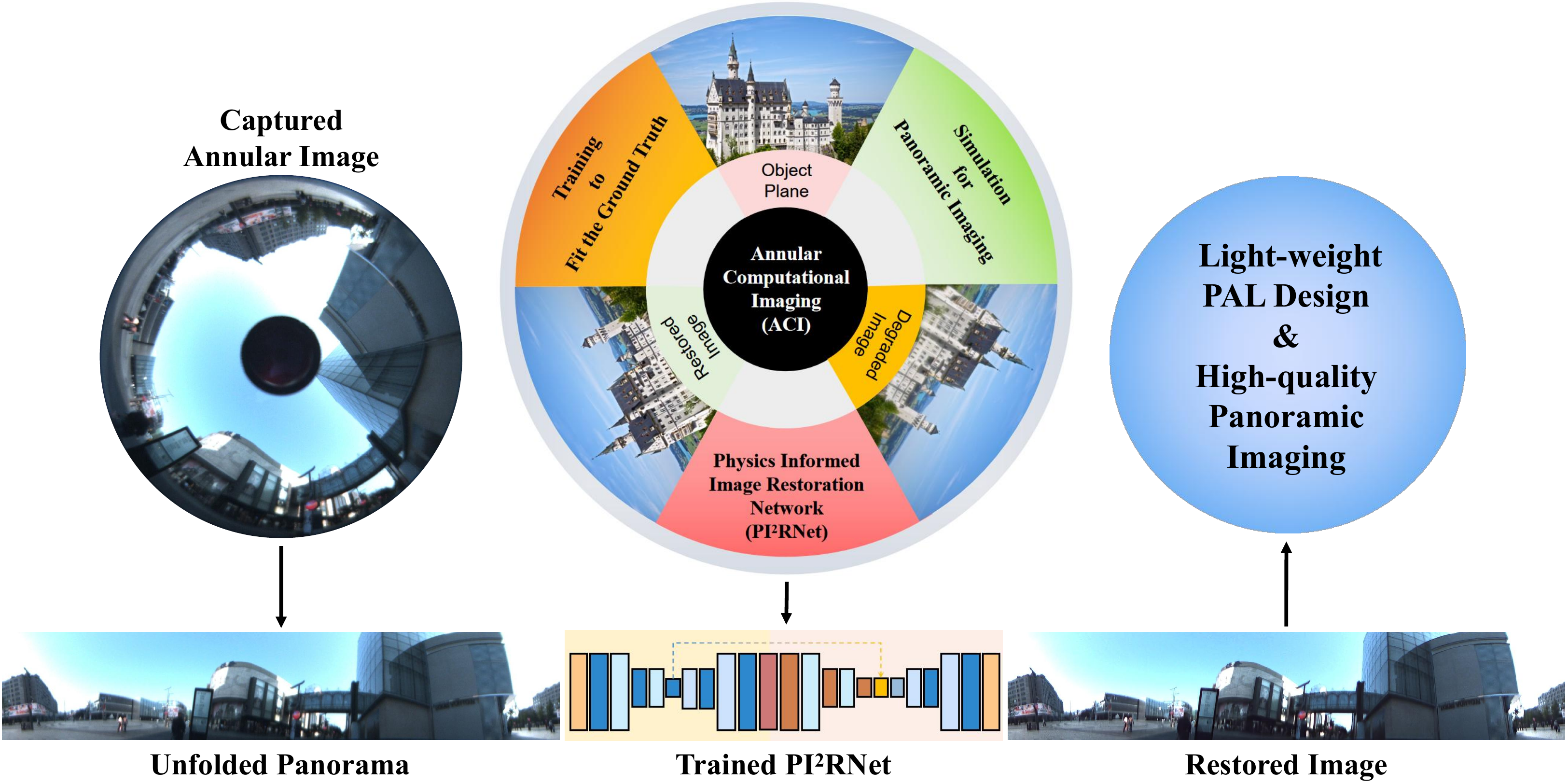}
\vskip-2ex
\caption{The proposed ACI framework. The ``annular'' closes the loop when the restored image approximate the high-quality sharp image. The proposed {\rm PI$^{2}$RNet} is trained on synthetic degraded dataset through optical simulation and restore the degraded image captured by a PAL with compact design. We map the annular image into unfolded one for the similar data distribution of perspective image by which our network is trained with. In this way, the restored image from {\rm PI$^{2}$RNet} enables light-weight PAL design with high imaging quality. For more detailed description please refer to Sec.~\ref{sec:Framework}.}
\label{fig:framework}
\vskip-3ex
\end{figure*}

\subsection{Imaging Simulation under Optical Aberration}
Data-hungry supervised learning methods require hundreds of degraded-sharp image pairs for training. Although manual acquisition methods~\cite{2020Replacing,peng2019learned} which capture the same scene with high-end cameras and simple lens can produce real degraded-sharp image pairs, the complex alignment and shooting process make this method restrict to specific case and difficult to transfer to other different lenses. 

Consequently, imaging simulation~\cite{Chenshiqi,hu2021image,sun2021end} becomes a preferable choice by generating synthetic aberration datasets. The key point is to calculate the PSFs under optical aberration, which are often achieved by calibration, raytracing or wave-based methods. PSFs calibration is often assisted by low-rank decomposition~\cite{xue2022deep, yanny2022deep, yanny2020miniscope3d} to improve the sparsely calibrated PSFs in LSV system. Raytracing is a direct method for PSF calculation. Shih \textit{et al.}~\cite{shih2012image} calculate the PSFs of given lens based on the design parameters by combining raytracing with calibration, while Sun \textit{et al.}~\cite{sun2021end} and Chen \textit{et al.}~\cite{Chenshiqi} produce synthetic data pairs through raytracing-based imaging simulation. In contrast, wave-based methods~\cite{peng2016diffractive,sitzmann2018end,hu2021image} are more flexible and can generate multiple distributions of PSFs, whose core part is to estimate the phase function on pupil plane. Zernike polynomials model is widely adopted to represent the phase under aberration~\cite{dun2020learned}, especially when the influence of aberration is complex~\cite{Xiang2010Effects,Xiang2017Aberrations}. 

Yet, due to the special catadioptric imaging mechanism of PAL, no approaches are explored for its imaging simulation. In addition, most of aforementioned pipelines rely on the lens design while ignoring the errors introduced by manufacture and assembly, making them highly depend on the low tolerances of the lens design. Hence, our simulation is set up for PAL in particular and takes into account the gap between the synthetic data and real panoramic imaging through wave-based PSF model.

\subsection{Panoramic Computer Vision Tasks}
Panoramic images for vision tasks are often obtained by image mosaic or panoramic cameras. Multi-view image mosaic is widely adopted for panoramic imaging with 360$^\circ$ FoV. Patil \textit{et al.}~\cite{2005People} and Deng \textit{et al.}~\cite{2017Object} perform panoramic object detection task with multiple perspective and fisheye lenses respectively. Lin \textit{et al.}~\cite{lin2018pvo} propose panoramic visual odometry for a multi-camera system with two fisheye lenses, while Seok \textit{et al.} design Rovo~\cite{seok2019rovo} and Omnislam~\cite{won2020omnislam} for wide-baseline multi-camera systems in subsequent years. 

Recently, computer vision tasks with 360$^\circ$ FoV panoramic images from PAL show advantages. Yang \textit{et al.} propose panoramic annular semantic segmentation (PASS)~\cite{yang2019pass} which is valuable for surrounding sensing~\cite{yang2019can} based on PAL. Chen \textit{et al.}~\cite{2019PALVO} design the PALVO which applies PAL to visual odometry by modifying the camera model and specially designed initialization process. Fang \textit{et al.}~\cite{fang2020panoramic} employ PAL for visual place recognition, leveraging panoramas for omnidirectional perception with FoV up to 360$^\circ$. Shi \textit{et al.}~\cite{shi2022panoflow} explore 360$^\circ$ optical flow estimation based on the cyclicity of panoramic images.

The robust and efficient performance of the panoramic vision algorithm depends on high-quality images captured by panoramic lens, while the cost and portability of the equipment is crucial if applied in mass-produced mobile terminals. In comparison to multi-camera systems, PAL is a more compact panoramic solution capable of imaging with FoV of 360$^\circ$. Nevertheless, current designs~\cite{cheng2016design, gao2021design} require additional lenses to optimize image quality, otherwise the produced images are with severe aberrations. We propose the ACI framework to recover the images for low-cost PAL designs based on computational methods and make a sweet pot between the trade-off of the high-quality imaging and light-weight lens design. We hope that PAL design adopting ACI framework can maintain both tiny volume and high imaging quality, benefiting the panoramic computer vision community.

\section{Overview}
\label{sec:Framework}

\begin{figure*}[!t]
\centering
\includegraphics[width=1\linewidth]{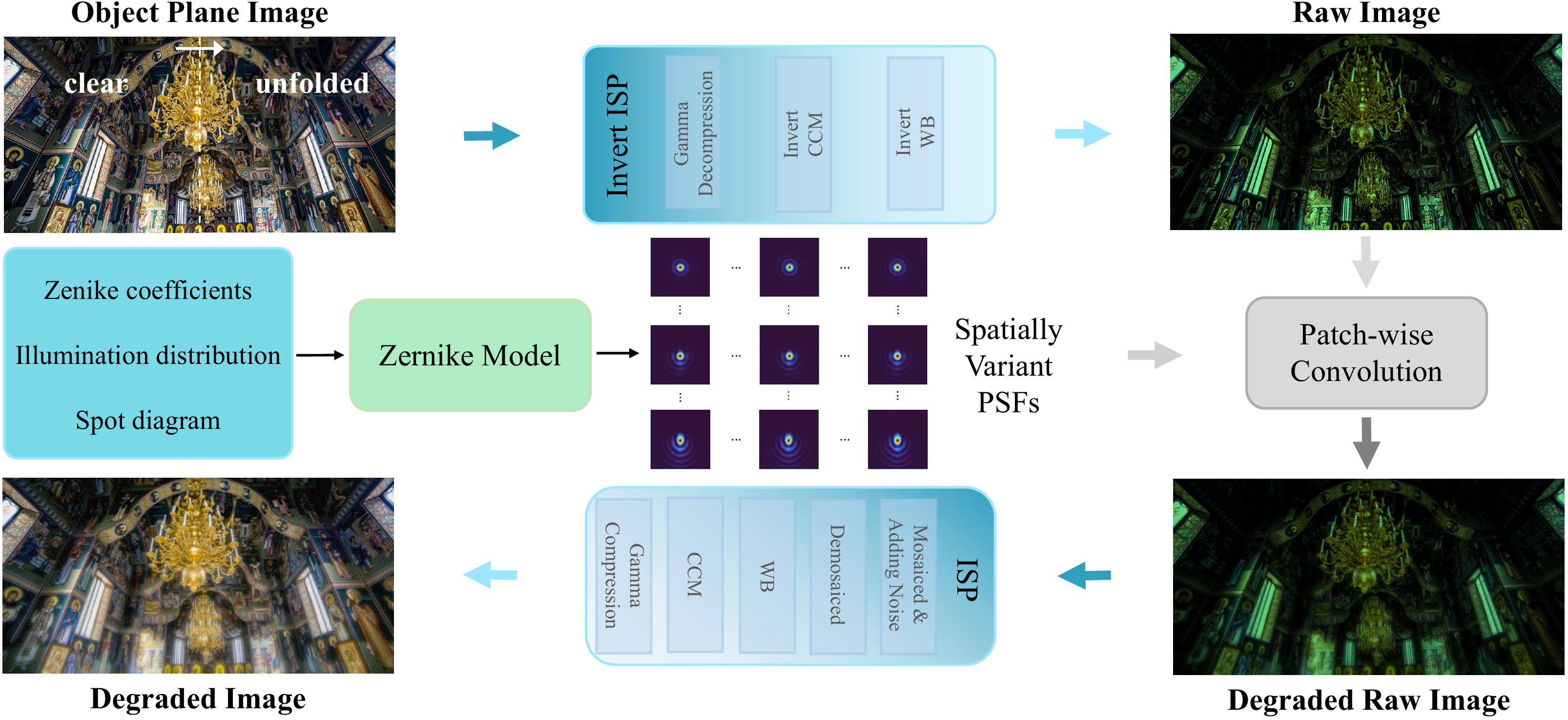}
\vskip-2ex
\caption{Simulation pipeline for panoramic annular imaging. A perspective clear image on object plane is downsampled and upsampled asymmetrically to simulate the PAL unfolding process. The object plane image then degrades to aberration image based on aberration distribution on unfolded PAL image plane. The image signal processing (ISP) contains mosaiced, adding noise, demosaiced, white balance (WB), color correction matrix (CCM), and gamma compression while the Invert ISP is composed of gamma decompression, invert color correction matrix (Invert CCM), and invert white balance (Invert WB). The patch-wise convolution convolves the raw image after invert ISP with spatially variant PSFs from Zernike Model considering the Zernike coefficients, illumination distribution and spot diagram from the given PAL.}
\label{fig:simulation_pipeline}
\vskip-3ex
\end{figure*}

For PAL, compact design with fewer lenses means sacrifice for image quality. Tiny volume and high image quality are contradictory in optical design. The goal of our work is to correct the image degradation of simple designed PAL through computational methods, enabling 360$^\circ$ FoV panoramic lens design with more compact structure, higher image quality, and lower manufacturing costs. The captured degraded annular image is unfolded and then fed to the CNN to recover a panoramic image without degradation (as shown in Fig.~\ref{fig:framework}). Since perspective high quality images are publicly available, the PAL image is processed in unfolded type because it has similar distribution with public datasets. The annular flow chart in Fig.~\ref{fig:framework} shows the proposed ACI framework. The degraded image is synthesized from high-quality perspective images through asymmetrical sampling and optical simulation. Then we train the {\rm PI$^{2}$RNet} to reconstruct the image from simulated degradation, making it fit the ground truth well. Due to the similar degradation of synthesized dataset and real captured unfolded panorama, the {\rm PI$^{2}$RNet} can produce clear unfolded panoramic images. Detailed descriptions of the simulation pipeline and {\rm PI$^{2}$RNet} are given in Sec.~\ref{sec:simulation} and Sec.~\ref{sec:network} respectively.

\section{Simulation for Panoramic Annular Imaging}
\label{sec:simulation}
For panoramic image restoration, a fatal problem is the lack of datasets, especially for the PAL annular image. However, the unfolded PAL image has similar data distributions with perspective images which is publicly available. It is reasonable to approximate the perspective images as unfolded PAL images and simulate corresponding degradation on them after considering the uneven sampling induced by unfolding. By this means, we can enjoy the benefits of large amount of the existing datasets. 

The simulation pipeline for panoramic annular imaging is depicted in Fig.~\ref{fig:simulation_pipeline}, which maps clear perspective images to images with degradation characteristics of the given PAL. Our process follows the principle of real imaging equipment,\ie, the modulation of optical system is on the raw image~\cite{2019Unprocessing} while the sensor presents the image we are familiar with through image signal processing (ISP) systems. It can be expressed as:
\begin{equation}
\label{eq:conv}
A = \Gamma(R \otimes K + N).
\end{equation}
The key point of the simulation is to compute the PSF $K$ of the optical system. At last, aberration image $A$ is produced by ISP $\Gamma(\cdot)$ after the convolution $\otimes$ between the raw image $R$ and PSF with the addition of sensor’s shot and read noise $N$.

The rest of this section is organized as follows: We first discuss the calculation of phase function on pupil plane and focal intensity on image plane in a general way. Then, the solutions to the particularities of PAL are depicted. Finally, we review the whole pipeline which completes the simulation for panoramic annular imaging.

\subsection{Phase Function on Pupil Plane }
Without loss of generality, we select general imaging system with circular pupil as a example, and firstly only discuss PSF calculation under a certain fixed FoV $\theta_0$ and wavelength $\lambda_0$. Assuming that a point light source is on the object plane, the intensity distribution on the image plane after propagation is the PSF. 

According to Fourier optics, the pupil function on pupil plane affected by aberration needs to be described first, which counts for PSF calculation. In this case, it’s unnecessary to discuss the specific types of aberrations, because the effect of them is all to deviate the wavefront at the pupil from the ideal sphere. The wave aberration $W(x,y)$ is applied to represent the optical path difference between the two surfaces, so the phase function $\Phi(x,y)$ on the pupil plane is expressed as:
\begin{equation}
\label{eq:phase}
\Phi(x,y) = k_0W(x,y),
\end{equation}
where $k_0=\frac{2\pi}{\lambda_0}$ denotes the wave number. 

Zernike polynomials are adopted as a mathematical description of optical wavefronts propagating through circular pupils~\cite{2011Zernike}. $\Phi(x,y)$  can therefore be described by Zernike circle polynomials $\mathcal{W}(\rho,\theta)$ in polar coordinates as:
\begin{equation}
\label{eq:zernike}
\mathcal{W}(\rho,\theta) = \sum_{n,m} {C^m_n}{Z^m_n}(\rho,\theta),
\end{equation}
where $C$ denotes Zernike coefficients and $Z$ refers to polynomials. The combination of different $m$ and $n$ represents different orders. For the specific expressions of each item please refer to ~\cite{V1994Zernike}. The above polar expression of $(\rho,\theta)$ can be rewritten in rectangular coordinates of $(x,y)$ by coordinate transformation ($x=\rho\cos\theta, y=\rho\sin\theta$).

With the phase function on the pupil, pupil function $\mathcal{P}(x,y)$ can be written by:
\begin{equation}
\label{eq:pupil}
\mathcal{P}(x,y) = P(x,y)e^{\mathrm{i}\mathcal{W}(x,y)},
\end{equation}
where $P(x,y)$ is circ function in our case. 

\subsection{Focal Intensity on Image Plane}
In such imaging system, the light field propagating from the back surface to the image plane satisfies Fresnel diffraction theory and the amplitude can be calculated by scalar diffraction integral~\cite{huggins2007introduction}. Before integral, we distinguish the coordinates of the image plane and the pupil plane. Let the coordinate of the image plane be $(x,y)$ and that of the pupil plane be $(x',y')$. Therefore, the amplitude of the image plane $E(x,y)$ can be expressed as:
\begin{equation}
\label{eq:integral}
E(x,y) = \frac{E_0}{{\lambda_0}d}\iint\mathcal{P}(x',y')e^{-\mathrm{i}\frac{2\pi}{{\lambda_0}d}(x'x + y'y)}dx'dy',
\end{equation}
where $E_0$ is a constant amplitude related to illumination, $d$ denotes the distance from the pupil plane to the image plane and $\lambda_0$ is the selected wavelength mentioned above. 

The above discussion is carried out at a certain FoV $\theta_0$ and wavelength $\lambda_0$. In fact, the calculated $E$ on image plane is a function of FoV $\theta$ and wavelength $\lambda$. More generally, we write equation \eqref{eq:integral} as:
\begin{equation}
\label{eq:integrallambda}
E(x,y,\theta,\lambda) = \frac{E_0}{{\lambda}d}\iint{\mathcal{P}(x',y',\theta,\lambda)e^{-\mathrm{i}\frac{2\pi}{{\lambda}d}(x'x + y'y)}dx'dy'}.
\end{equation}
The dependence of pupil function $\mathcal{P}(x',y',\theta,\lambda) = P(x',y')e^{\mathrm{i} \mathcal{W}(x',y',\theta,\lambda)}$ on FoV and wavelength is reflected in wave aberration $\mathcal{W}(x',y',\theta,\lambda)$ which has diverse distributions under different FoVs and wavelengths. 

Finally, the focal intensity on the image plane is:
\begin{equation}
\label{eq:intensity}
I(x,y,\theta,\lambda) = {|E(x,y,\theta,\lambda)|^2}.
\end{equation}
Under our hypothetical point light source condition, the above intensity is the PSF distribution $K(x,y,\theta,\lambda)$ required.

\subsection{Adaptation to Unfolded PAL Images}
\label{sec:adapt}

Equations~\eqref{eq:phase}-~\eqref{eq:intensity} provide the calculation approach for PSFs in general imaging system. In this section, the adaptation to PAL will be discussed. Since most of the available high-quality images are perspective, our simulation for panoramic annular imaging is aimed at the unfolded PAL image plane. Following previous works~\cite{scaramuzza2006flexible}, we adopt the unit sphere model to describe the projection relationship for the catadioptric system. A unit sphere which takes the camera as its origin is included with the Taylor polynomial to fit the parametric model of PAL. We calibrate the coefficients of the Taylor polynomial for PAL through the omnicalib toolbox~\cite{scaramuzza2006toolbox}. Then, we apply the camera model to project the original image onto the unit sphere and use the EquiRectangular Projection (ERP) to produce an equirectangular panorama unfolded image. 

For an annular PAL image, the patches of the inner circle with fewer pixels will be upsampled during unfolding, resulting in the degradation of image quality. This process can be expressed as upsampling with different scale factors which should be considered in simulation. The annular image is unfolded based on a certain FoV $\theta_0$, and the width of the unfolded image is the circumference of the circle at $\theta_0$. For a panoramic camera whose vertical field of view contains 90$^\circ$, we take $\theta_0$ as 90$^\circ$. For cameras that not included, we take the $\theta_0$ angle closest to the equator on the unit sphere~\cite{scaramuzza2006toolbox}. The scale factor $S_{\theta}$ of any FoV $\theta$ is calculated by:

\begin{equation}
\label{eq:scale factor}
S_{\theta} = \frac{2\pi r_{\theta_0}}{2\pi r_{\theta}},
\end{equation}
where $r_{\theta_0}$ is the radius of $\theta_0$ and $r_{\theta}$ is the radius of $\theta$. We divide a perspective image into patches longitudinally and downsample each patch with corresponding scale factor $S_{\theta}$. After upsampling these patches back to their original resolutions, the degradation of unfolding is simulated on the perspective image. 

\begin{figure}[!t]
\centering
\includegraphics[width=1.0\linewidth]{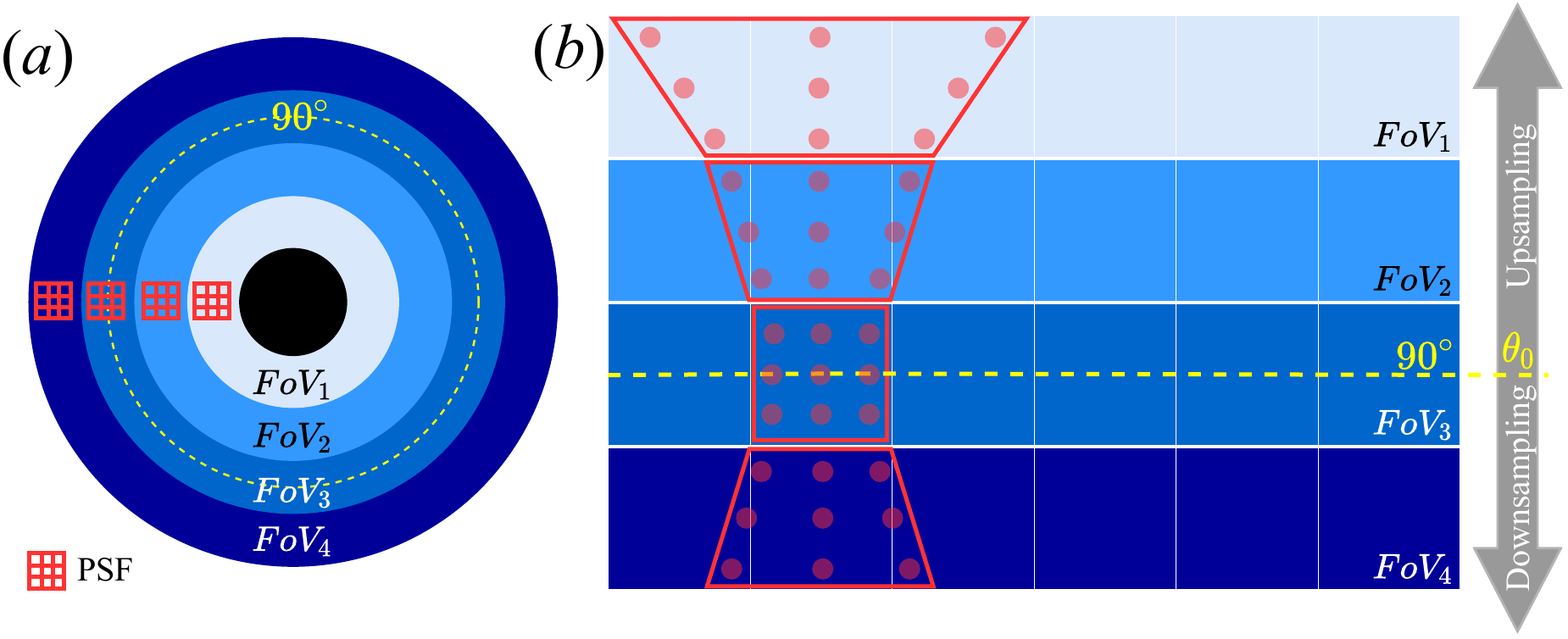}
\vskip-2.5ex
\caption{Schematic of how FoVs and PSFs distribute on the annular image plane and unfolded image plane of the PAL, respectively. (a): Annular image plane. (b) Unfolded image plane. Note that the original images with FoVs smaller than and larger than $\theta_0$ are up- and down-sampled to produce the unfolded image. The PSF kernels are distorted differently according to their corresponding FoVs. We only sample FoVs in a sparse way and use $3\times3$ kernels for a simple demonstration.}
\label{fig:PSF-dist}
\end{figure}

Moreover, we consider the distribution of PSFs and the geometric deformation introduced by projection in the unfolding process. As shown in Fig.~\ref{fig:PSF-dist}, PSFs annularly distribute with FoVs along the radial direction on the original PAL image plane. On the unfolded image plane, PSFs distribute along the transverse stripes instead. Therefore, each row of pixels in the image corresponds to one PSF under its FoV. In addition, the geometric deformation of the PSFs in the ERP format is linearly related to the introduced scale factor $S_{\theta}$ of each FoV. The squire PSF kernels on annular image plane in Fig.~\ref{fig:PSF-dist}(a) is distorted to a trapezoidal one on unfolded image plane in Fig.~\ref{fig:PSF-dist}(b) based on distribution of $S_{\theta}$ with respect to FoVs ($S_{\theta}$ varies in vertical but keeps fixed in horizontal). In this way, based on the kernel size on annular image computed from spot diagram, we can obtain the mapped size of PSF kernels on unfolded image plane, adapting the PSFs calculation to unfolded PAL images.

\begin{figure}[t]
\centering
\includegraphics[width=1.0\linewidth]{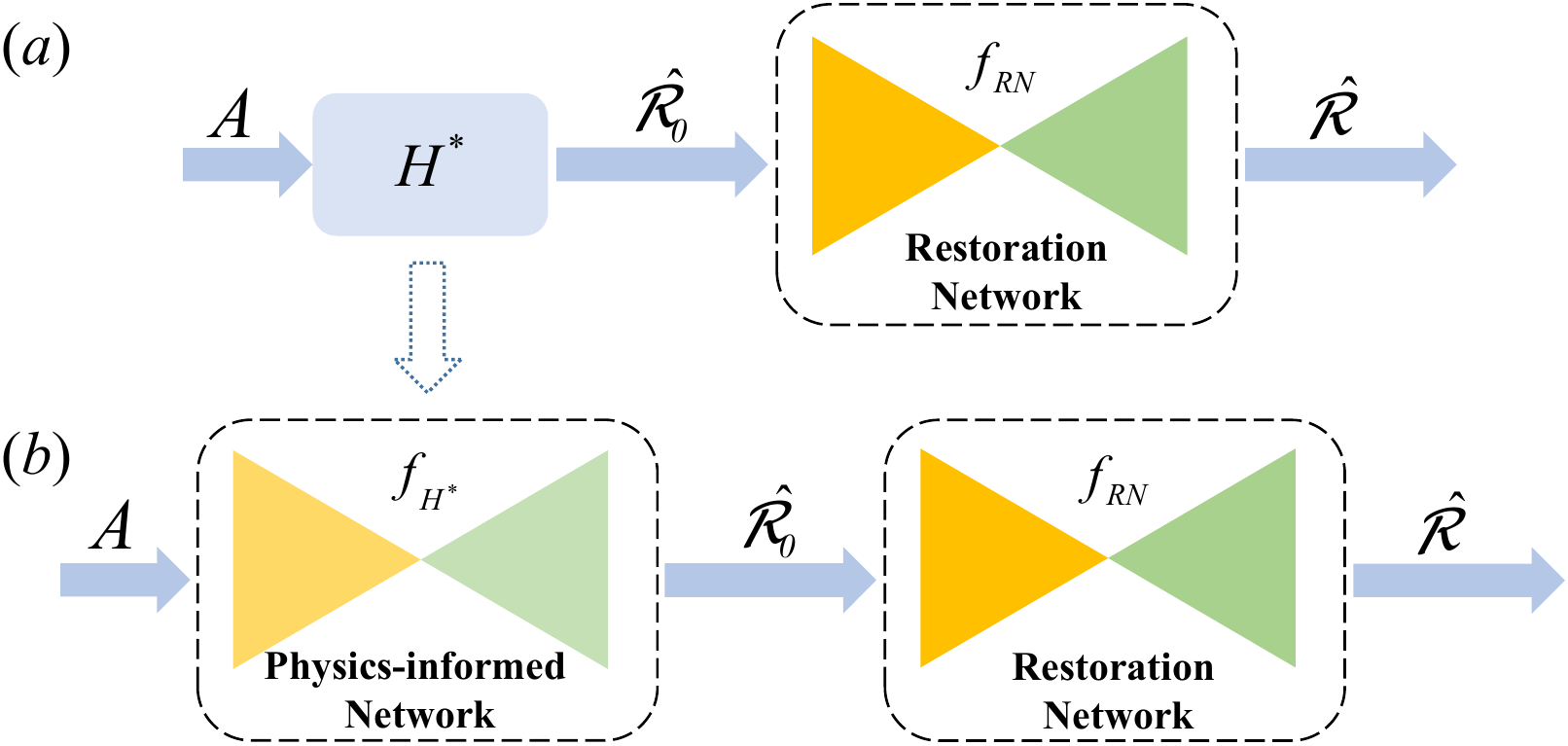}
\vskip-2.0ex
\caption{The proposed single-pass physics-informed engine. (a) The engine in general CI. (b) The adaptation to our {\rm PI$^{2}$RNet}. $A$ is the original image to be recovered and $\hat{\mathcal{R}_0}={H^*A}$ is a single approximant of the final restored image $\hat{\mathcal{R}}$. }
\label{fig:sp}
\vskip-2.5ex
\end{figure}

\begin{figure*}[!t]
\centering
\includegraphics[width=1.0\linewidth]{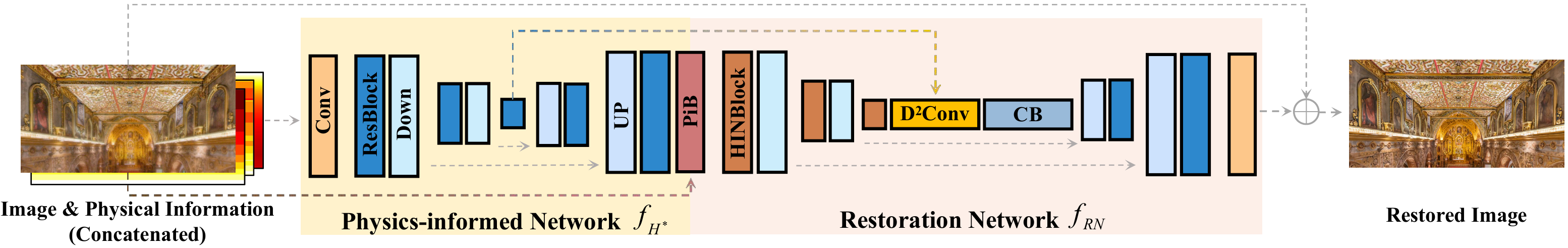}
\vskip-1.5ex
\caption{An overview of the proposed Physics Informed Image Restoration Network ({\rm PI$^{2}$RNet}). With physical information (height distribution of PSFs, width distribution of PSFs and scale factor distribution) concatenated with degraded image, Physics-informed Network is applied to make an approximate restoration while extracting robust degradation characteristics from physical information. Then the characteristics are integrated to the Restoration Network through Physics-informed Bridge (PiB, introduced in Sec.~\ref{sec:pib}). ``Down'': downsample, ``Up'': upsample. ``D$^{2}$Conv'' (Dynamic Deformable Convolution, introduced in Sec.~\ref{sec:d2conv}) is utilized based on the encoded physical feature from Physics-informed Network $f_{H^*}$ and the ``CB'' (Context Block~\cite{MengChang2020SpatialAdaptiveNF}) helps keep the global context of features at the bottleneck of Restoration Network $f_{RN}$. }
\label{fig:network}
\vskip-3ex
\end{figure*}

\subsection{Simulation Pipeline}
For the simulation pipeline described in equation~\eqref{eq:conv}, we need a more rigorous expression. The calculated PSF $K(x,y,\theta,\lambda)$ on unfolded PAL image plane is a function of FoV $\theta$ and wavelength $\lambda$. Under a certain FoV, equation~\eqref{eq:conv} can be expressed as the combined action of all wavelengths within corresponding band:
\begin{equation}
\label{eq:conv2}
A(x,y,\theta) = \Gamma(\int{R(x,y,\theta)\otimes K(x,y,\theta,\lambda)d\lambda}),
\end{equation}
where the process of adding noise is included in the ISP $\Gamma(\cdot)$. $A(x,y,\theta)$ is the image patch of degraded image under FoV $\theta$ and $R(x,y,\theta)$ is the corresponding patch on raw image. We sample and divide the raw image into patches according to the FoV distributions of unfolded PAL images described in Sec.~\ref{sec:adapt}. The image patch is convolved with PSF under $\theta$ and then merged to form the whole degraded image.

The Zernike coefficients of $\mathcal{W}(x',y',\theta,\lambda)$ can be calculated by $Zemax^\circledR$ during lens design. This process requires to sample different FoVs and wavelengths within the imaging range. In addition, the illumination distribution and the spot diagram of the lens are also considered to determine $E_0$ and the size of PSF kernels respectively.

The invert ISP in Fig. 3 includes gamma decompression, invert color correction matrix, and invert white balance while the ISP contains mosaiced, adding noise, demosaiced, white balance, color correction matrix and gamma compression. For details of these image processing operations, we refer readers to the work of Brooks \textit{et al.}~\cite{2019Unprocessing}.

In conclusion, we propose a pipeline to synthesise degraded images with degradation characteristics of unfolded PAL images, providing a data generation approach for panoramic image restoration. Instead of raytracing to directly calculate fixed-distribution PSFs~\cite{Chenshiqi} or completely setting random coefficients for universal datasets~\cite{hu2021image}, our pipeline takes the aberration distribution of designed lens as a reference and randomly adjusts Zernike coefficients within a controllable range to produce additional data pairs. Although the PSFs produced by raytracing is more accurate, the manufacture and assembly errors are ignored. Our solution is more like a data argumentation that closes the synthetic-to-real gap and has loose-tolerance requirements. 

\section{Phyiscs Informed Image Restoration Network}
\label{sec:network}

Based on the physical model in Sec.~\ref{sec:simulation}, we first adapt the single-pass physics-informed engine~\cite{Barbastathis:19} (Sec.~\ref{sec:singlepass}) to our work. In Sec.~\ref{sec:architecture}, we introduce the whole network architecture composed of double U-nets named Physics-informed Network and Restoration Network. The special convolution kernel designed for spatially varying PSFs and the bridge to inform the Restoration Network the physical information from Physics-informed Network are depicted in Sec.~\ref{sec:d2conv} and Sec.~\ref{sec:pib} respectively. 

\subsection{Single-pass Physics-informed Engine}
\label{sec:singlepass}
 In CI problem where the reconstructed image $\hat{\mathcal{R}}$ is predicted from input $A$, Barbastathis \textit{et al.}~\cite{Barbastathis:19} adopt the idea of physics-informed learning, which helps solve inverse problems efficiently through data and physical model, to incorporate physical priors into deep learning, and propose the single-pass physics-informed engine demonstrated in Fig.~\ref{fig:sp} (a). Consistency with ~\cite{Barbastathis:19}, we use the operator $H$ to represent the forward model from $\hat{\mathcal{R}}$ to $A$. $H^*$ is the approximate inverse operator of $H$ so that the engine can be modeled as:

\begin{equation}
\label{eq:sp}
\hat{\mathcal{R}} = f_{RN}({H^{*}A}; \Theta),
\end{equation}
where $f_{RN}(\cdot)$ denotes the Restoration Network and $\hat{\mathcal{R}_0}={H^{*}A}$ in Fig.~\ref{fig:sp} is a single approximant of the restored image. $\Theta$ is learnable parameters for Restoration Network. In our case, the forward operator $H$ is the process described in equation~\eqref{eq:conv2}, in which the information of PSFs: $K(x,y,\theta,\lambda)$ is available from simulation pipeline. 

However, solutions to the approximate inverse operator $H^*$ are no longer convincing in our case because of the difference between simulated PSFs and the real ones. Consequently, we design the novel Physics-informed Network $f_{H^*}(\cdot)$ as a substitute for operator $H^*$ and adapt the engine to our work. As shown in Fig.~\ref{fig:sp} (b), the proposed Physics-informed Network makes an approximate restoration $\hat{\mathcal{R}_0}$. The delicate design of Physics-informed Network has potential for better integration of physical priors.

\subsection{Network Architecture}
\label{sec:architecture}

We propose the Phyiscs Informed Image Restoration Network to address the complex degradation introduced by spatially-variant PSFs and uneven upsampling. As shown in the Fig.~\ref{fig:network}, our model consists of two major parts: Physics-informed Network  $f_{H^*}(\cdot)$ and the Restoration Network $f_{RN}(\cdot)$. 

\begin{figure}[!t]
\centering
\includegraphics[width=0.8\linewidth]{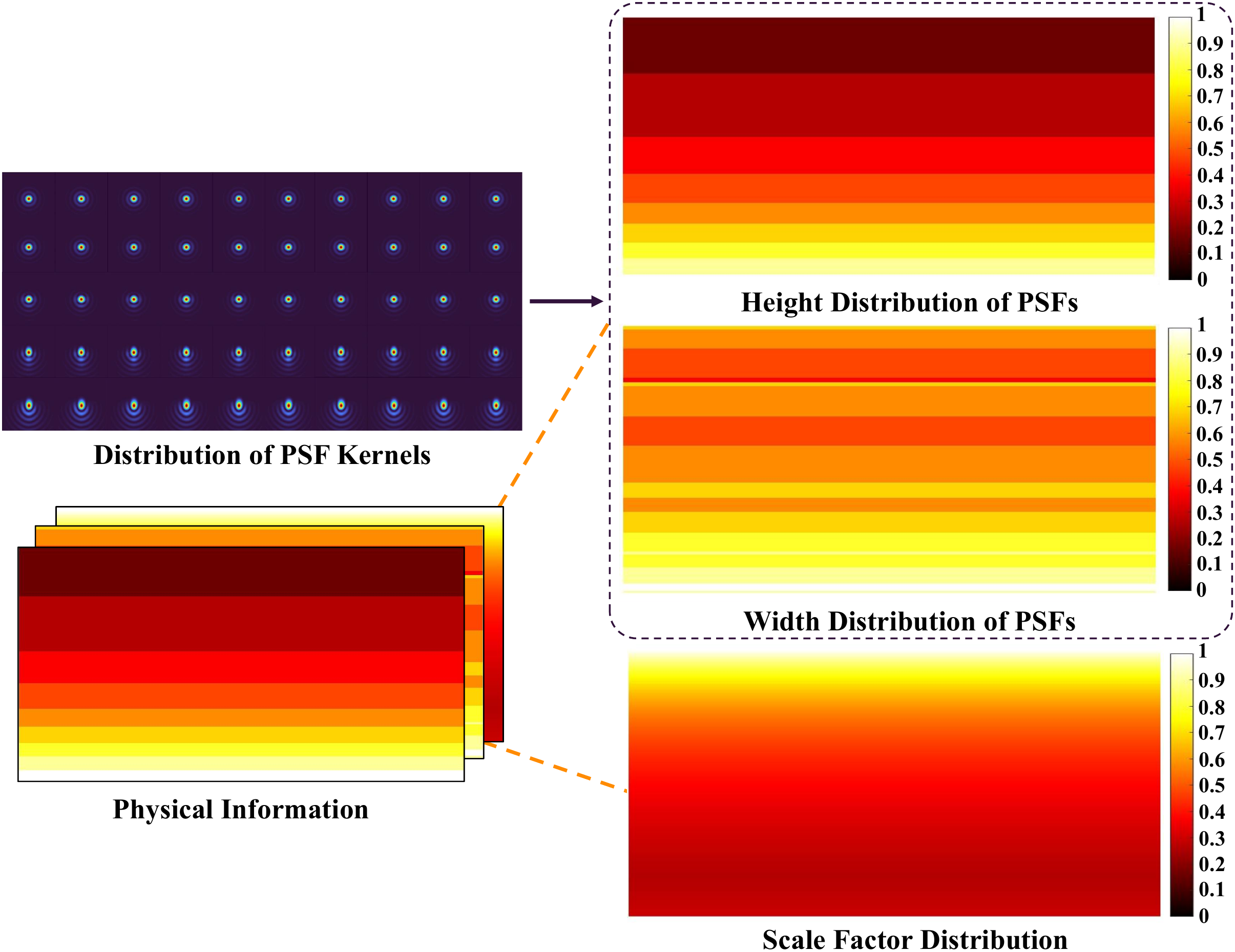}
\caption{Composition of physical information. We extract PSF kernel size (height and width distribution of PSFs) from the distribution of PSF kernels. Concatenated with the scale factor distribution, the physical information indicates the degradation level in different patches.}
\label{fig:physicalinfo}
\vskip-2ex
\end{figure}

The degraded image and physical information are firstly fed into Physics-informed Network to make an approximate restoration. As Fig.~\ref{fig:physicalinfo} shows, our physical information consists of two parts: The distributions of PSF kernel sizes (height and width), and scale factors that indicates the degradation information of different spatial regions of the unfolded image captured by particular PAL. To enable $f_{H^*}(\cdot)$ to extract degradation characteristics from physical features and transfer them from $f_{H^*}(\cdot)$ to $f_{RN}(\cdot)$, we propose the Physics-informed Bridge (PiB) in the bottleneck of two encoder-decoder network.

The encoder and decoder of Restoration Network is composed of HINBlocks and ResBlocks respectively adopted from HINet~\cite{chen2021hinet} which proposes half-instance normalization for image restoration. The Down operation is a 4$\times$4 convolution with stride 2 and the Up operation is a 2$\times$2 transpose convolution with stride 2. Consider that the PSFs which generate aberration are spatially-variant in both content and shape, we apply Dynamic Deformable Convolution (D$^{2}$Conv) to feature maps between encoder and decoder based on encoded feature map from $f_{H^*}(\cdot)$. In addition, to maintain the global context information and enlarge the receptive filed, Context Block (CB)~\cite{MengChang2020SpatialAdaptiveNF} is introduced which also constructs multi-scale information and makes network robust to noise.

\subsection{Dynamic Deformable Convolution}
\label{sec:d2conv}

For PAL designs with very few lenses, the aberration distribution varies greatly with FoVs, causing the shape and content of PSFs spatially-variant. The traditional convolution operation overlooks this variation and applies spatially-fixed convolution kernel to pixels across the whole image. Since the image degradation is relevant to the spatial distribution of PSFs based on equation~\eqref{eq:conv2}, we should redesign our convolution kernel to tackle the challenge of spatially-variant image degradation.

Deformable Convolution (DConv)~\cite{dai2017deformable} proposes a novel convolution format, in which the kernel shape changes with the position of pixels, while the content of the kernel is fixed for all locations across the feature map. Coincidentally, the Filter Adaptive Convolution (FAC) layer~\cite{zhou2019spatio} is able to generate different kernel content for pixels with different spatial coordinates. To handle the spatially-variant degradation of images from PAL, we combine kernel-shape-variant convolution (\ie~DConv) with kernel-content-variant convolution (\ie~FAC) and propose our Dynamic Deformable Convolution.

\begin{figure}[!t]
\centering
\includegraphics[width=1.0\linewidth]{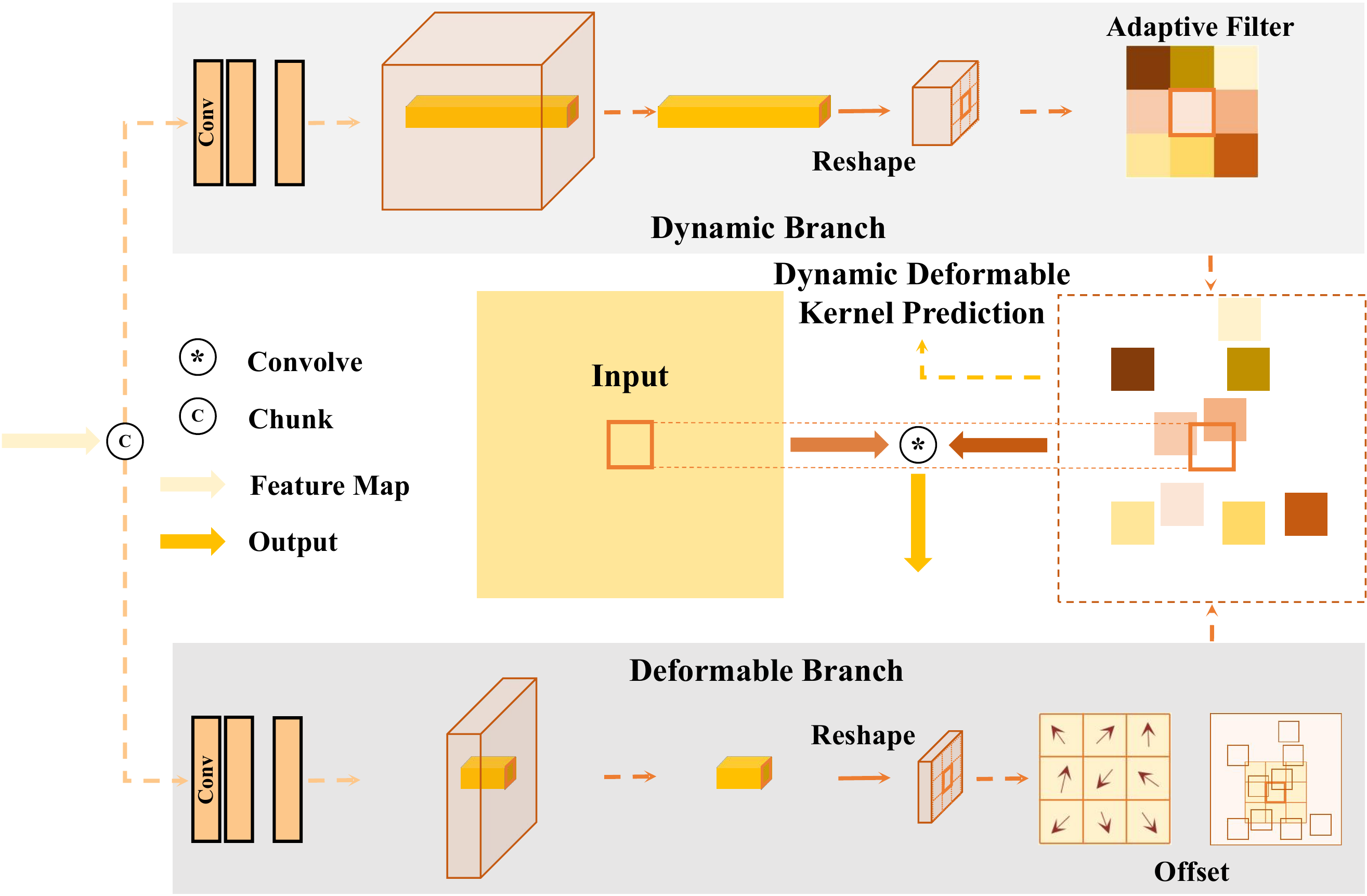}
\vskip-1.5ex
\caption{The proposed Dynamic Deformable Convolution (D$^{2}$Conv). The feature map containing physical information is chunked into two branches to make a Dynamic Deformable Kernel Prediction (D$^{2}$KP), where the dynamic branch produces an adaptive filter for every pixel and the deformable branch predicts the offset for it. Then, the input image (or feature) is convolved with the kernel.}
\label{fig:d2conv}
\end{figure}

As is illustrated from Fig.~\ref{fig:d2conv}, the feature map is firstly chunked into two branches to make a Dynamic Deformable Kernel Prediction (D$^{2}$KP). The dynamic branch produces an adaptive filter for every pixel and the deformable branch predicts the offset of each coordinate to decide the filter shape for it. Then, the input image (or feature maps) is convolved with the filter. When the feature map contains physical information, the filter will be adaptive to each pixel in terms of shape and content based on the degradation model.

\subsection{Physics-informed Bridge}
\label{sec:pib}
\begin{figure}[!t]
\centering
\includegraphics[width=1\linewidth]{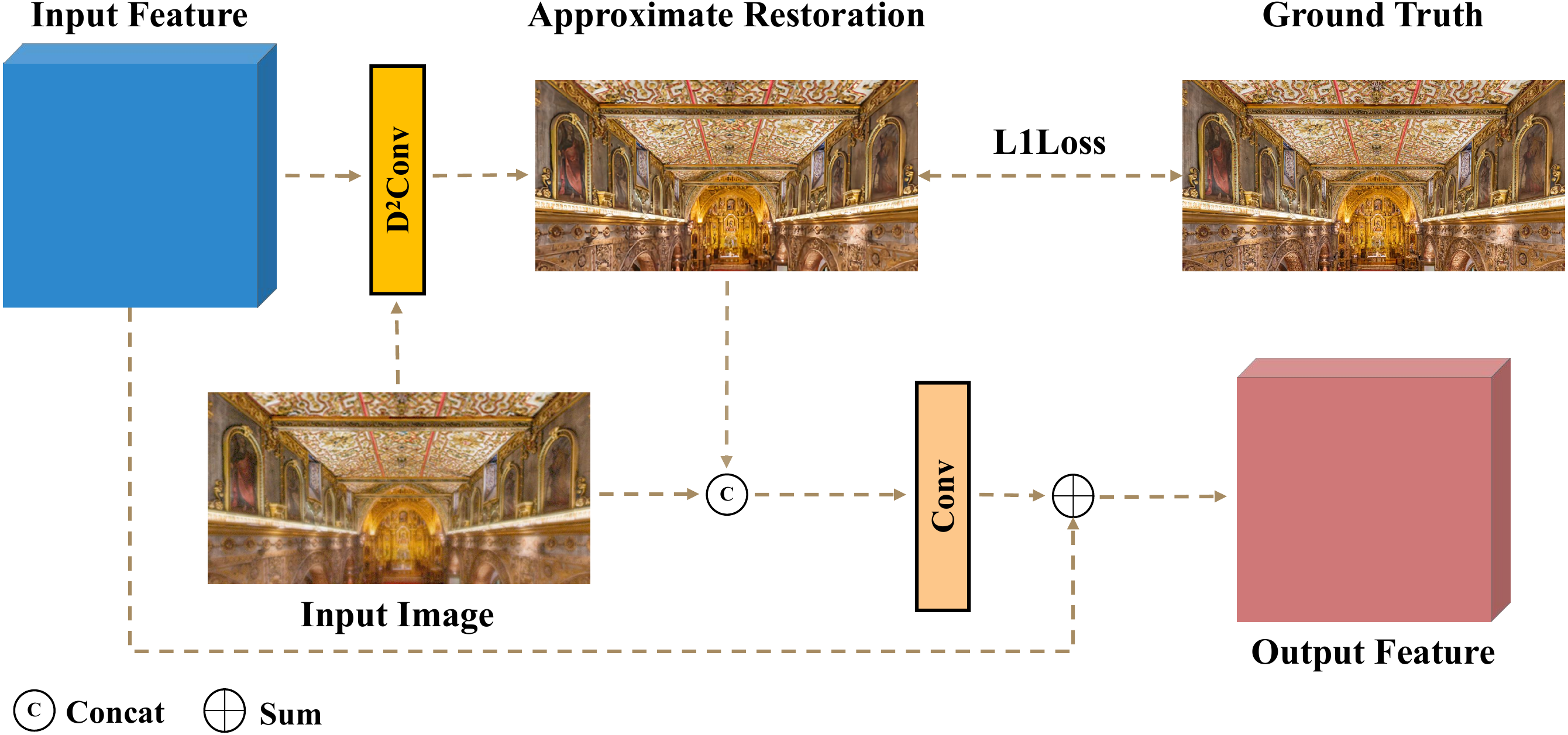}
\vskip-1.5ex
\caption{The proposed Physics-informed Bridge (PiB). Through proposed D$^{2}$Conv, we perform D$^{2}$KP to enable $f_{H^*}(\cdot)$ to reconstruct feature of complex degradation. The decoded feature from $f_{H^*}(\cdot)$ (\ie~input feature) is also integrated with image feature into the Restoration Network. PiB is the bridge to transfer the physical information between $f_{H^*}(\cdot)$ and $f_{RN}(\cdot)$. }
\label{fig:PIB}
\vskip-3ex
\end{figure}

A simple U-net to represent $H^*$ may cause the loss of physical information and the coarse restoration process is not connected with the convolution model in degradation, bringing poor performance (depicted in Sec.~\ref{sec:ablation}). Inspired by the convolution model of image degradation, Kernel Prediction (KP)~\cite{vogels2018denoising} is often applied to predict the dynamic kernel weights and biases for image restoration, but restricted to fixed kernel shape in different pixels. Thus, we replace the KP with our D$^{2}$KP to better model the inverse process of panoramic imaging. Consequently, we build a Physics-informed Bridge (illustrated in Fig.~\ref{fig:PIB}) to equip $f_{H^*}$ with physical significance and pass on modeled physical feature to $f_{RN}(\cdot)$.

As Fig.~\ref{fig:PIB} shows, we make D$^{2}$KP first with the decoded feature map from $f_{H^*}(\cdot)$. In PiB, the input image is applied D$^{2}$Conv with the kernel for an approximate restoration supervised by the ground truth with L1Loss to ensure that $f_{H^*}(\cdot)$ is trained to reconstruct feature of complex distributions of PSFs. Then, we fuse the physical features with the feature maps from input image and produce the output feature for $f_{RN}(\cdot)$. In this way, the physical model of image restoration in single-pass physics-informed engine is replaced with our Phyiscs-informed Network and its physical priors are transferred to the Restoration Network.  

\section{Experiments}

\subsection{Setup}

\subsubsection{Optical System}

\begin{figure}[!t]
\centering
\includegraphics[width=1\linewidth]{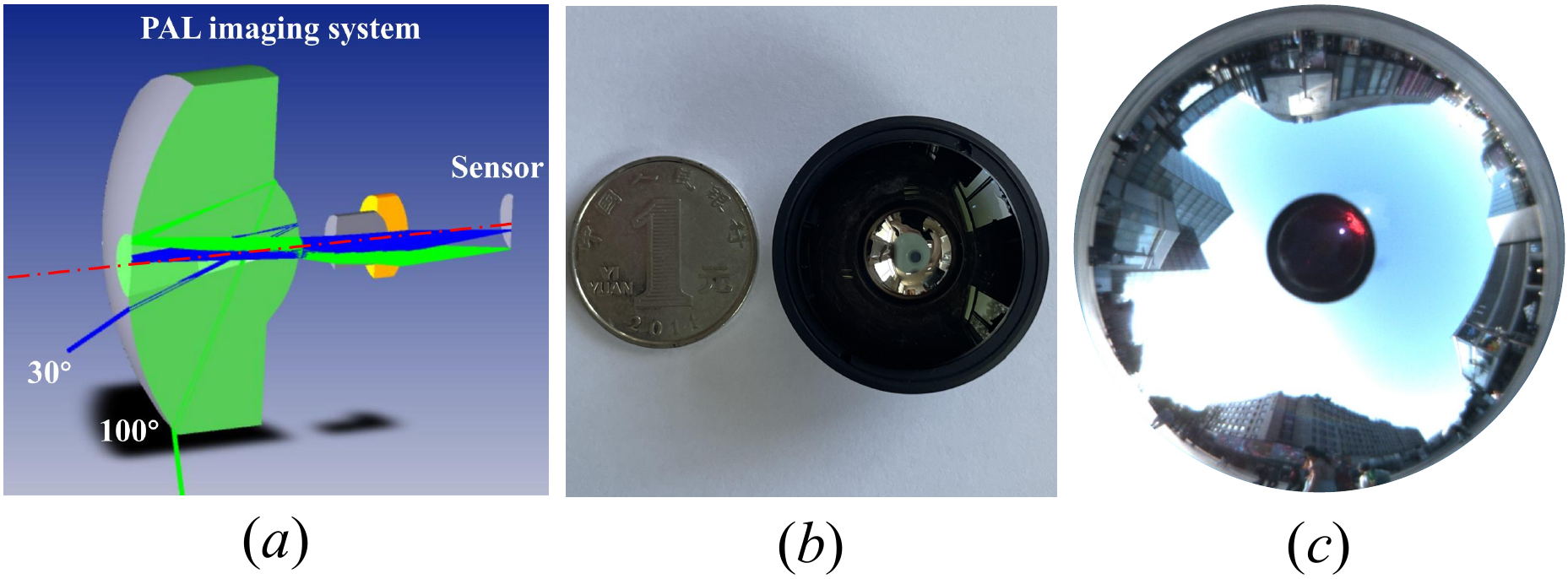}
\caption{The optical system for the experiments of ACI. (a) The light path diagram of the selected PAL. (b) Real PAL system with tiny size. (c) The captured degraded image.}
\label{fig:pal}
\vskip2.5ex
\end{figure}

Our optical system for experiments on ACI is a PAL system consisting of three spherical lenses which is thin and light. Our system applies the prototype of a PAL head of one lens and a relay lens group composed of several lenses. Considering the crucial catadioptric structure of the PAL head for 360$^\circ$ panorama, we can only reduce the number of lenses of the relay lens group for more compact system, \ie~2 lenses in our case. The FoV of the system is 360°$\times$(30°-100°). Reasonable focal power distribution of lenses according to the power distribution principle can greatly simplify the number of lenses, promoting its application under volume and space constraints. The light path diagram of selected PAL is shown in Fig.~\ref{fig:pal}(a). The blue rays are rays path of the 30° half FoV. Green rays are rays with a maximum half FoV of 100°. The number of lenses of this PAL is reduced to less than 30$\%$ of traditional one, leading a relatively deficient imaging quality with aberration.

Fig.~\ref{fig:pal}(b) shows the adopted PAL system, the maximum optical lens diameter of our device is comparable to the size of a coin, enabling panoramic CV tasks for space-constrained scenarios. The MV-SUA133GC camera with resolution of 1024 $\times$ 1280 and pixel size of 4$\mu{m}$ is applied as the sensor considering the comparable target plane size with our PAL. Due to the reduction of the number of lenses, the optical aberration of the large FoV is hard to correct through optical design, resulting in the degradation of the captured image shown in Fig.~\ref{fig:pal}(c). The proposed ACI framework will alleviate the drawback of relatively low imaging quality while keeping the tiny size. 

\subsubsection{Data Preparation}
\label{sec:divpano}
We generate degraded images through applying simulation pipeline on DIV2K~\cite{Agustsson_2017_CVPR_Workshops}. The illumination distribution, spot diagram and Zernike coefficients (keep the first 37 polynomials as a common practice~\cite{V1994Zernike, 2011Zernike}) under each FoV and wavelength are calculated by $Zemax^\circledR$. For each wavelength, the FoVs are sampled in the range of [30$^\circ$,100$^\circ$] with step of 0.7$^\circ$ for 101 related PSFs. We produce 31 groups of such PSFs (PSFs from 101 FoVs in each group) from wavelengths of 400$nm$ to 700$nm$ in 10$nm$ intervals. Then we translate them to red, green, and blue channels according to the wave response of the sensor, respectively. Considering that the panorama is unfolded to 288 $\times$ 1504, we crop and downsample the images of DIV2K into 512 $\times$ 1024 and produce degraded images through simulation pipeline (depicted in Sec.~\ref{sec:simulation}) based on aforementioned 101 PSFs in RGB channels.

To improve the generalization ability of our model to complex degradation distribution, we randomly adjust each Zernike coefficient at all FoVs within $\pm$25$\%$ to produce image pairs with diverse aberration distributions. For each FoV $\theta$, a random number of the uniform distribution $r_{i,\theta}$ within $[-0.25, 0.25]$ is generated, and coefficient $Z_{i,\theta}$ is then replaced by $(1+r_{i,\theta})Z_{i,\theta}$.

\begin{figure}[!t]
\centering
\includegraphics[width=1.0\linewidth]{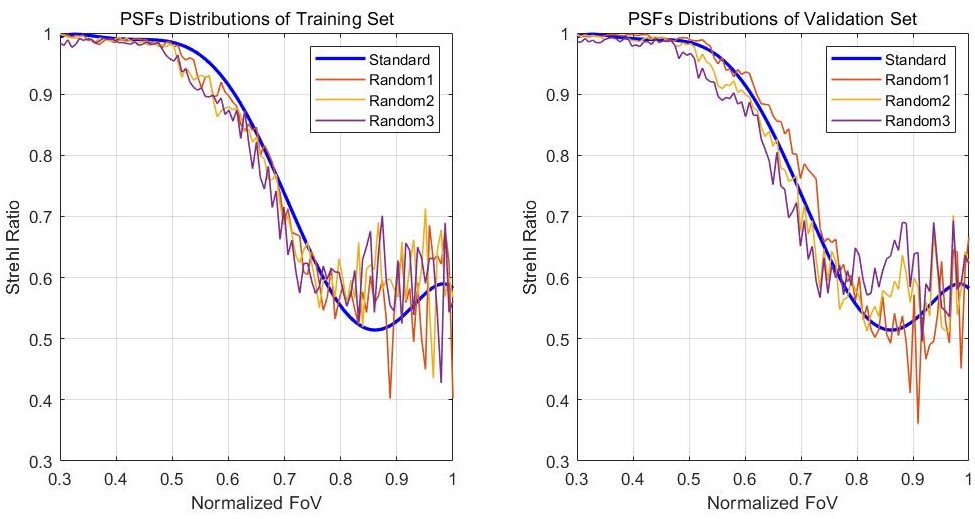}
\vskip-1.5ex
\caption{The PSFs distributions of our DIVPano dataset. Strehl ratio calculated in unfolded PAL image plane is used to describe the distribution. The different random distributions can help bridge the synthetic-to-real gap and validate the generalization ability of the restoration models. }
\label{fig:datadist}
\end{figure}

By this means, we propose the DIVPano dataset composed of image pairs under different aberration distributions from random adjustment. To validate the generalization ability of image restoration approaches, the aberration distributions in validation set is different from those in training set. We use Strehl ratio~\cite{1982Strehl}, which can measure the energy diffusion of PSFs, to describe the aberration distributions under different FoVs. As is shown in Fig.~\ref{fig:datadist}, centering around a standard distribution, training set consists of 3 deviated PSFs distributions while validation set is composed of other 3 different ones. The ground truth for training set and validation set are divided in the same way as DIV2K while the testing images are captured by the chosen PAL.

\subsubsection{Training Details}
The DIVPano dataset and the physical information introduced in Sec.~\ref{sec:architecture} are used to train our {\rm PI$^{2}$RNet}. The approximate restoration and final restored image are both supervised by ground truth $\mathcal{G}$ with L1Loss: 

\begin{equation}
\label{eq:loss}
\mathcal{L} = {\lambda}{\parallel}\hat{\mathcal{R}_0}-\mathcal{G}{\parallel}+{\parallel}\hat{\mathcal{R}}-\mathcal{G}{\parallel},
\end{equation}where $\lambda$ is the loss weight and empirically set to 0.5. We also conduct experiments on the choice of $\lambda$ in Sec.~\ref{sec:ablation}.

The network is trained with Adam optimizer on a single NVIDIA GeForce RTX 3090. We also use rotation and flip as data argumentation. We set the initial learning rate to 2$\times$10$^{-4}$ and apply cosine annealing strategy~\cite{loshchilov2016sgdr} with a minimum learning rate of 1$\times$10$^{-7}$. Finally, we train our model on random crops of 256$\times$256 image patch with a batch size of 8 for 1.4$\times$10$^5$ iterations. 

\subsection{Ablation Study}
\label{sec:ablation}
We conduct sufficient ablation studies on the components of physics-informed engine (Tab.\ref{tab:ablation_single}), comparison of different design in the D$^{2}$Conv module (Tab.\ref{tab:ablation_d2}), different context components (Tab.\ref{tab:ablation_context}), and different data distributions (Tab.~\ref{tab:ablation_random}) of our DIVPano dataset. Peak Signal-to-Noise Ratio (PSNR) and Structural Similarity (SSIM)~\cite{wang2004image} are employed as our evaluation metrics for all the experiments. 

Ablations on single-pass physics-informed engine are shown in Tab~\ref{tab:ablation_single}. Our baseline (row 1) is a cascade structure of two U-nets which has considerable parameters with our network. We first concatenate the physical information with our input image into the baseline (row 2). Physical information can inform network the degradation distributions and improves PSNR by 0.67 dB and SSIM by 0.71$\%$. Then we adopt the single-pass engine to make an approximate restoration but without our PiB. The drop in both metrics (\ie~ 0.9 dB in PSNR and 4.6$\%$ in SSIM, row 3 and 4) indicates that replacing the physical model of $H^*$ with a single U-net is ineffective. However, PiB is able to guide the former network to extract features of dynamic deformable kernels and inform the latter network the physical feature, contributing to an improvement of 0.3 dB in PSNR and 0.34$\%$ in SSIM (row 3 and 6). Even replacing the D$^{2}$KP in PiB with KP, we can achieve better results compared to that without PiB (\ie~1.15 dB in PSNR and 4.89$\%$ in SSIM, row 4 and 5) or single-pass engine (0.25 dB in PSNR and 0.29$\%$ in SSIM, row 3 and 5). 

\begin{table}[t]
\caption{\textbf{Ablations on Single-pass physics-informed engine.}}
  \centering
  \resizebox{0.45\textwidth}{!}{

  \renewcommand\arraystretch{1.4}{{
  \begin{tabular}{cccc}
            \bottomrule[0.15em]
            Experiment & Variations & PSNR$(dB)$ & SSIM$(\%)$\\
             
            \hline
            Baseline & - & 27.59 & 88.52\\
            
            \hline
            \multirow{2}{*}{Physical Information} & \textit{w/o} & 27.59 & 88.52\\
            & \underline{\textit{with}} & 28.26 & 89.23\\
            
            \hline
            \multirow{3}{*}{Physics-informed Bridge} & \textit{w/o} & 27.36 & 84.63\\
            & KP & 28.51 & 89.52\\
            & \underline{D$^{2}$KP} & \textbf{28.56} & \textbf{89.57}\\
            
            \bottomrule[0.15em]
     \end{tabular}}}}
     \vspace{-1em}
\label{tab:ablation_single}
\end{table}

Ablations on D$^{2}$Conv Module are shown in Tab~\ref{table:abd2conv}. The baseline is the network with single-pass engine and D$^{2}$KP PiB evaluated above. We replace the D$^{2}$Conv in the bottleneck of Restoration Network with DConv~\cite{dai2017deformable} and FAC~\cite{zhou2019spatio} respectively to evaluate the effectiveness of different convolution kernels. The network with D$^{2}$Conv stands out with improving PSNR by 0.14 dB and SSIM by 0.13$\%$ compared to baseline (row 1 and 4). D$^{2}$Conv considers both kernel-shape-variant convolution and kernel-content-variant convolution, which is consistent with the degradation process. 

\begin{table}[t]
  \caption{\textbf{Ablations on D$^{2}$Conv Module.}}
  \label{tab:ablation_d2}
  \centering
  \resizebox{0.45\textwidth}{!}{
  \renewcommand\arraystretch{1.45}{\setlength{\tabcolsep}{8.25mm}{\begin{tabular}{cccc} 
            \bottomrule[0.15em]
            Method & PSNR$(dB)$ & SSIM$(\%)$ \\
            
            \hline
            
            Baseline & 28.56 & 89.57 \\
            DConv~\cite{dai2017deformable} & 28.59 & 89.65 \\
            FAC~\cite{zhou2019spatio} & 28.47 & 89.42 \\
            D$^{2}$Conv & \textbf{28.70} & \textbf{89.70} \\

            \bottomrule[0.15em]
         \end{tabular}}}}
         \vspace{-1.0em}
\label{table:abd2conv}
\end{table}

\begin{table}[t]
  \caption{\textbf{Ablations on Global Context.}}
  \label{tab:ablation_context}
  \centering
  \resizebox{0.45\textwidth}{!}{
  \renewcommand\arraystretch{1.725}{\setlength{\tabcolsep}{5mm}{\begin{tabular}{cccc} 
            \bottomrule[0.15em]
            \multicolumn{2}{c}{\underline{ \quad Context Components \quad }} & \multirow{2}{*}{PSNR$(dB)$} & \multirow{2}{*}{SSIM$(\%)$} \\
            D$^{2}$Conv & CB~\cite{MengChang2020SpatialAdaptiveNF} & & \\
            
            \hline
            
            - & - & 28.56 & 89.57 \\
            \Checkmark & - & 28.70 & 89.70 \\
            - & \Checkmark & 28.53 & 89.59 \\
            \Checkmark & \Checkmark & \textbf{28.71} & \textbf{89.70} \\

            \bottomrule[0.15em]
         \end{tabular}}}}
         \vspace{-1.0em}
\label{table:abcontext}
\end{table}

\begin{table}[t]
  \caption{\textbf{Ablations on Random Distribution of Aberration.}}
  \label{tab:ablation_random}
  \centering
  \resizebox{0.45\textwidth}{!}{
  \renewcommand\arraystretch{1.4}{\setlength{\tabcolsep}{4.5mm}{\begin{tabular}{cccc}
            \bottomrule[0.15em]
            Distribution & Method & PSNR$(dB)$ & SSIM$(\%)$\\
             
            \hline
            \multirow{2}{*}{Standard} & SRN~\cite{tao2018scale} & 27.03 & 86.79 \\
            & Ours & 26.62 & 86.53 \\
            
            \hline
            \multirow{2}{*}{Random} & SRN~\cite{tao2018scale} & 28.26 & 88.58 \\
            & Ours & \textbf{28.71} & \textbf{89.70} \\
            
            \bottomrule[0.15em]
         \end{tabular}}}}
         \vspace{-1.0em}
\label{table:abrandom}
\end{table}

\begin{table}[t]
\begin{center}
\caption{\textbf{Ablations on loss weight $\lambda$.}}
\label{tab:lam}
  \resizebox{0.45\textwidth}{!}{
  \renewcommand\arraystretch{1.45}{\setlength{\tabcolsep}{8.25mm}{\begin{tabular}{cccc} 
            \bottomrule[0.15em]
            $\lambda$ & PSNR$(dB)$ & SSIM$(\%)$ \\
            
            \hline
            
            0.0 & 28.43 & 89.52 \\
           0.1 & 28.65 & 89.62 \\
            0.5 & \textbf{28.71} & \textbf{89.70} \\
            1.0 & 28.66&89.60 \\

            \bottomrule[0.15em]
         \end{tabular}}}}
\end{center}
\vspace{-2.5em}
\end{table}

\begin{figure}[!t]
\centering
\includegraphics[width=0.8\linewidth]{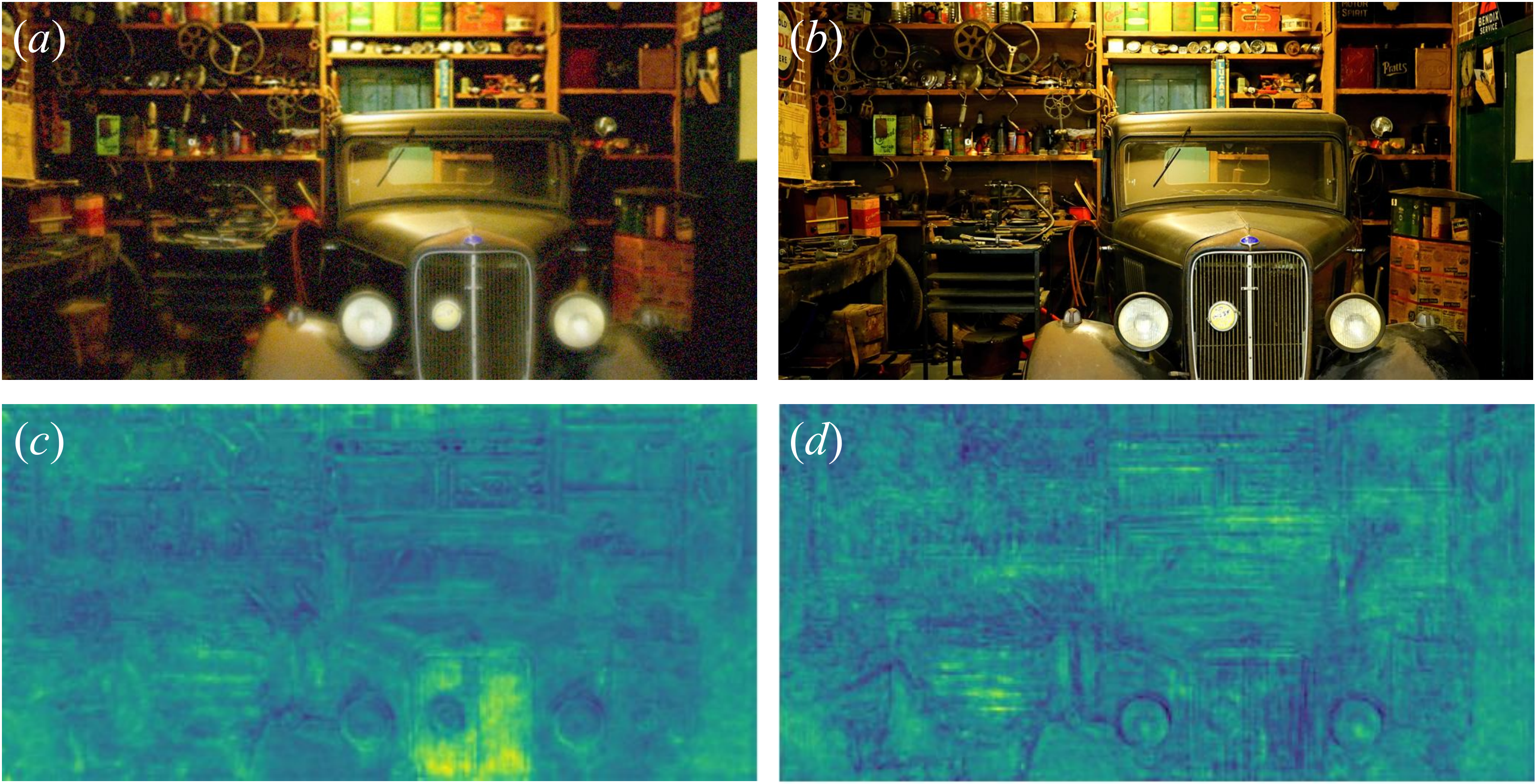}
\vskip-1.5ex
\caption{Effectiveness of D$^{2}$Conv and CB on global context. (a) and (b) are input degraded image and ground truth from DIVPano. The visualizations of feature maps without and with context components are shown in (c) and (d). }
\label{fig:CB}
\vskip-3ex
\end{figure}

Ablations on global context are shown in Tab~\ref{table:abcontext}. We also take the best model in row 6 of Tab~\ref{tab:ablation_single} as baseline and perform the ablation study. Except for the mentioned D$^{2}$Conv, CB is applied to enlarge the receptive field and helps keep the global context of features at the bottleneck of Restoration Network. Context components bring considerable improvement in metrics (\ie~0.15 dB in PSNR and 0.13$\%$ dB in SSIM, row 1 and 4) and make the deep features sharper and more even in global region (shown in Fig.~\ref{fig:CB}). 

\begin{figure*}[!t]
\centering
\includegraphics[width=1\linewidth]{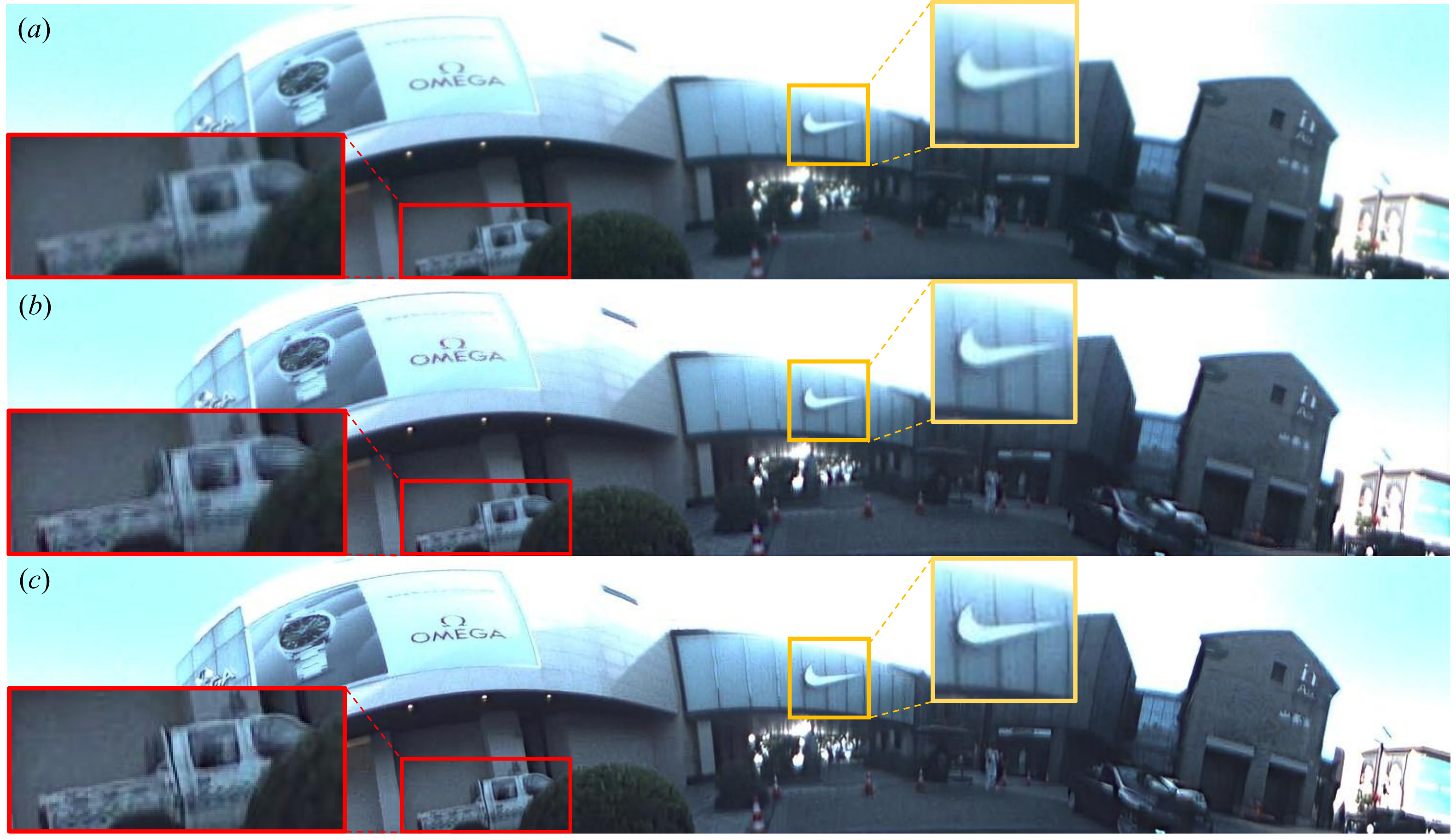}
\vskip-1.5ex
\caption{Random PSFs distributions \textit{v.s.} standard ones. (a) An unfolded image captured by the chosen PAL with aberration. (b) The restoration results under standard dataset. (c) The restoration results under random dataset. We Detailed patches in different FoVs are zoomed for best view. }
\label{fig:random}
\vskip-3ex
\end{figure*}

Ablations on random distributions of aberration for training set are shown in Tab~\ref{table:abrandom} and the corresponding results on a real captured unfolded PAL image are shown in Fig.~\ref{fig:random}. We train our {\rm PI$^{2}$RNet} and SRN~\cite{tao2018scale} (a classical network for image restoration) on training sets with standard PSFs distribution (blue curves in Fig.~\ref{fig:datadist}) and random distributions (other 3 curves around the standard one in Fig.~\ref{fig:datadist}). Random data distributions can improve the generalization ability of the model and enhance the performance under the synthetic-to-real gap. The quantitative results on validation set with other 3 random distributions illustrate the effectiveness of our dataset. We note that our method requires the random distributions for training a robust $f_{H^*}$ to learn degradation characteristics. Consequently, it performs more poorly than SRN when training with single standard distribution. In addition, the random distributions contribute to the better performance in our testing images. Compared with the degraded image in Fig.~\ref{fig:random}(a), the image in Fig.~\ref{fig:random}(b) restored by {\rm PI$^{2}$RNet} trained on standard distribution is slightly sharper but with more fake texture. Because of the single data distribution, the network overfits to the standard aberration characteristics and performs badly in real captured PAL images. Obviously, training with random data distributions brings a superior restoration, as shown in Fig.~\ref{fig:random}(c). 

Ablations on the loss weight $\lambda$ of the approximate restoration are shown in Tab~\ref{tab:lam}. When the $\lambda$ is set to 0, the modeling of degradation characteristics in $f_{H^*}$ becomes implicit, which is ineffective to the task and leads to the worse performance (28.71 dB to 28.43 in PSNR and 89.70 $\%$ to 89.52 $\%$ in SSIM). As a supervision for the learning of degradation characteristics, the loss function of the approximate restoration is essential which improves the PSNR by 0.22$\sim$0.28 and SSIM by 0.08$\%{\sim}$0.18$\%$. The optimal value for $\lambda$ is 0.5 with advantages of 0.05 dB in PSNR and 0.08$\%$ in SSIM.

\subsection{Comparisons with State-of-the-Art Methods}
We compare our proposed {\rm PI$^{2}$RNet} with state-of-the-art image restoration methods on both our DIVPano dataset and real unfolded PAL images. SRN~\cite{tao2018scale} and DeepRFT~\cite{mao2021deep} are selected to represent the multi-scale methods while DeepRFT processes images in frequency domain. KPN~\cite{mildenhall2018burst} is a classical network adopting KP which considers the spatially-variant degradation. In addition, we also choose latest NAFNet~\cite{chen2022simple} and HINet~\cite{chen2021hinet} because of their excellent performances in various image restoration tasks. We replicate these methods and retrain them on DIVPano with the default setting according to the corresponding papers. We keep batch size and training iterations the same as ours for all the methods for fair comparison.

Table~\ref{table:compare} shows the quantitative results (PSNR and SSIM) of selected methods on our validation set. The parameters and runtime for the inference of one unfolded panoramic image with resolution of $288 \times 1504$ tested on one single NVIDIA RTX 3090 are also illustrated. As mentioned in Sec.~\ref{sec:divpano}, the image degradation of our dataset varies greatly with FoVs, and to model the unknown aberration of the real lens, the aberration distribution is different between training set and validation set, placing a huge challenge for all the methods. Consequently, our {\rm PI$^{2}$RNet} which considers physical prior of panoramic imaging is superior to all the competitive methods. The physics-informed engine guides the network with the degradation distributions and proposed D$^{2}$Conv makes the recovery process conform to the degradation model. Our Physics-informed Network can learn robust characteristics of degradation based on physical information and degraded image from random distributions. In other words, the different degradation characteristics are first modeled and extracted by $f_{H^*}$.
As long as the degradation distributions of validation images approximates the theoretical design, the proposed $H^*$ promotes the network performance by strengthen the generalization of the model. All the proposed components and physical information help our method outperform the state-of-the-art methods by 0.11$\sim$2.38 dB in PSNR and 0.34$\sim$5.49$\%$ in SSIM on validation set of DIVPano, with slight increase in parameters and sacrifice of speed.

\begin{table}[t]
\begin{center}
\caption{\textbf{Comparison with state of the art on DIVPano.}}
\vskip-2ex
\label{table:compare}
  \resizebox{0.5\textwidth}{!}{
  \centering
  \renewcommand\arraystretch{1.45}{\setlength{\tabcolsep}{2mm}{\begin{tabular}{ccccc}
            \bottomrule[0.15em] 
            Method&PSNR$(dB)$&SSIM$(\%)$&Parameters(M)&Runtime(s)\\
            
            \hline
            SRN~\cite{tao2018scale} & 28.26 & 88.58&10.25&0.16 \\
            DeepRFT~\cite{mao2021deep} & 27.81 & 84.21&\textbf{09.60}&0.24 \\
            KPN~\cite{mildenhall2018burst} & 26.33 & 85.80&27.65&\textbf{0.10} \\
            NAFNet~\cite{chen2022simple} & 28.60 & 89.36&17.11&0.14 \\
            HINet~\cite{chen2021hinet} & 28.24 & 89.19&88.70&0.19 \\
            \midrule
            Ours & \textbf{28.71} & \textbf{89.70}&14.48&0.21 \\

            \bottomrule[0.15em]
         \end{tabular}}}}
\end{center}
\vspace{-2.5em}
\end{table}

\begin{figure*}[!t]
\centering
\includegraphics[width=1\linewidth]{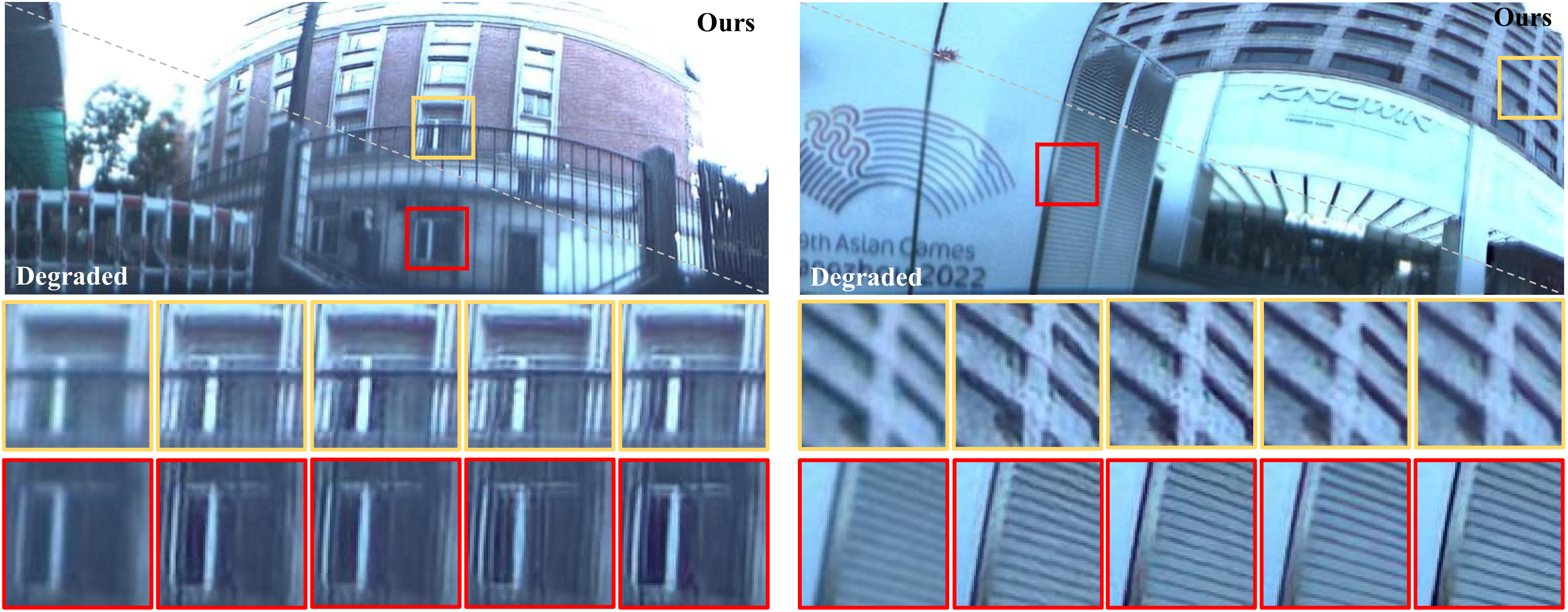}
\includegraphics[width=1\linewidth]{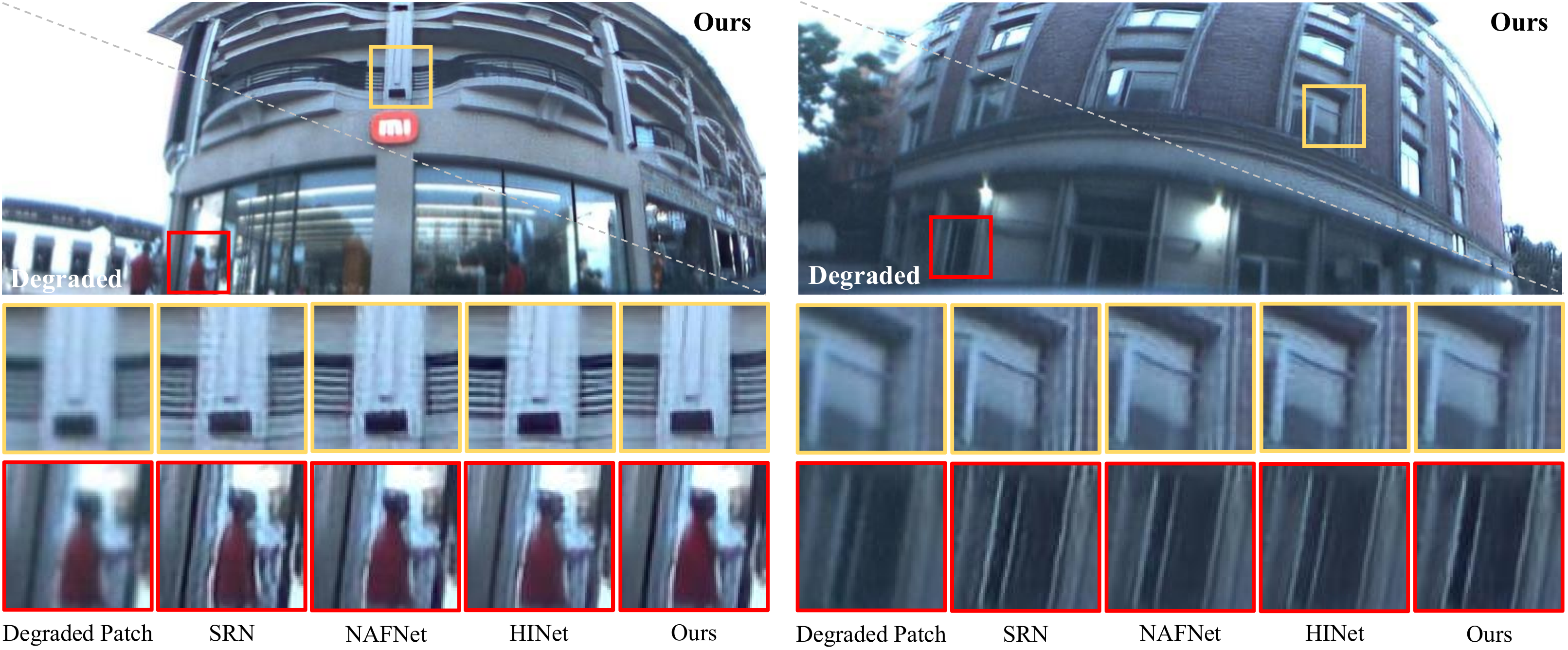}
\vskip-2ex
\caption{Qualitative results on real unfolded PAL image compared with state-of-the-art methods. According to Tab.~\ref{table:compare}, we compare our {\rm PI$^{2}$RNet} with other 3 most competitive methods, \ie~SRN, NAFNet and HINet. We unfold the image captured by selected PAL and shows a partial FoV of it at the top of each sub figure. The upper right part of this image is replaced with restoration result of our work. We highlight regions of details under different FoVs by chromatic boxes.}
\label{fig:qualitative}
\end{figure*}

We capture testing images with selected panoramic optical system to evaluate our model on real panoramic images. An unfolded PAL image is shown at the top of sub figure in Fig.~\ref{fig:qualitative} where the lower left part is the original degraded one and the upper right part is the image restored by our work. We select two image patches in each scene from different FoVs, and show the original image patches and restoration image patches from our method and other 3 best methods in Fig.~\ref{fig:qualitative}. With physical priors integrated, {\rm PI$^{2}$RNet} is a superior solution to the spatially-variant degradation of unfolded panorama. In the edge FoV regions (red boxes) with severer aberrations, our work has better aberration correction effectiveness in terms of higher contrast and sharper edges, while it can also produce convincing results with fewer artifacts and more realistic details in the regions with slight aberrations (yellow boxes).

The Physics-informed Network helps the network adapt to different degradation distributions without overfitting to synthetic data and prevents the restored image from fake textures, compared with the results from HINet. Comprehensively, based on the physical model in the image restoration, our proposed  {\rm PI$^{2}$RNet} dominates other state-of-the-art methods in restoration of unfolded PAL images.

\subsection{MTF Analysis}
\label{sec:mtf}
\begin{figure}[!t]
\centering
\includegraphics[width=1\linewidth]{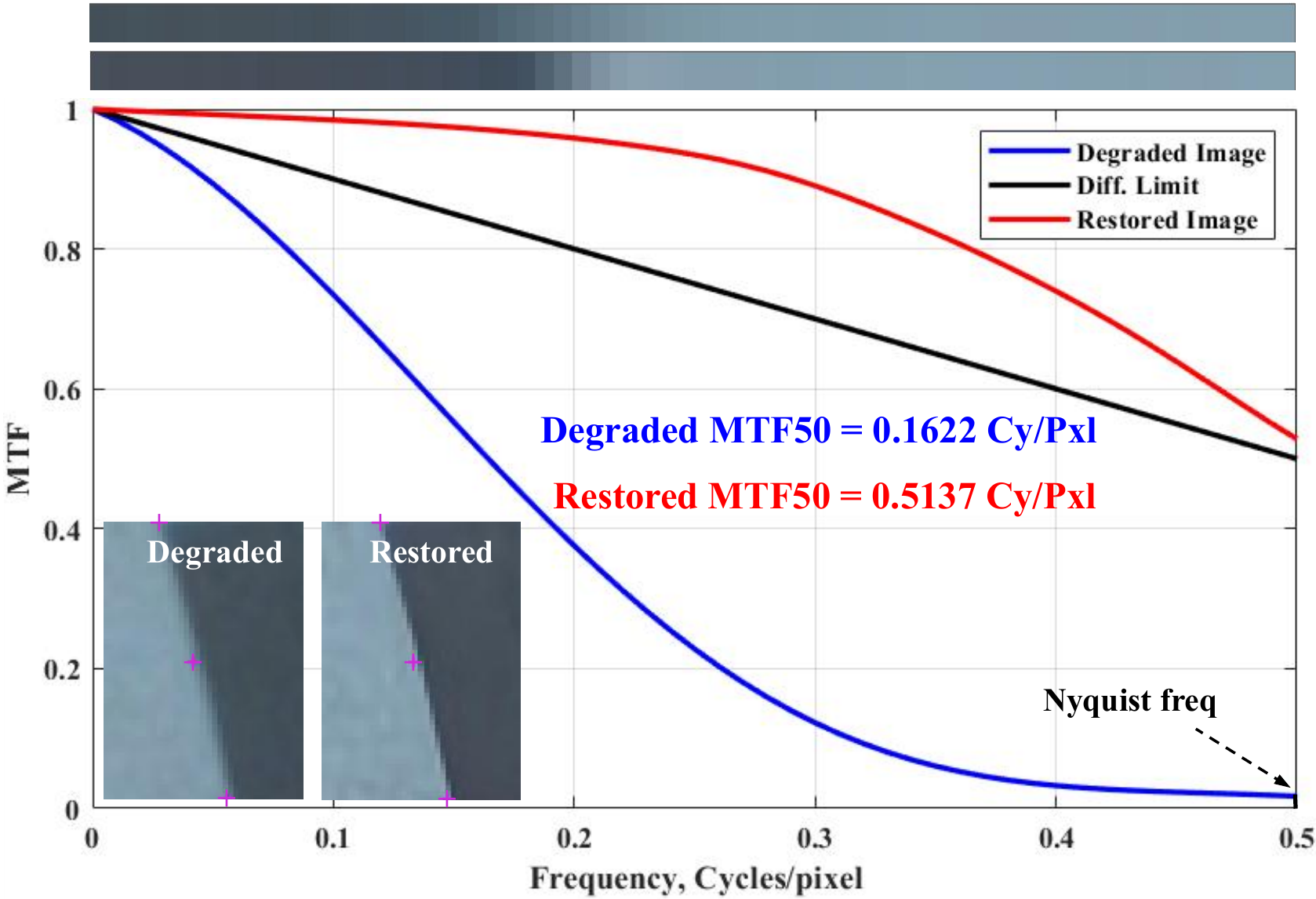}
\vskip-1ex
\caption{MTF analysis of unfolded PAL image. Average MTF curves of real unfolded PAL image and restored image by ACI are plotted along with the diffraction limit. The detected edges and corresponding image patches are shown on the top and lower left respectively. We also show the results of MTF50 of degraded image and restored image in the figure. }
\label{fig:mtf}
\vskip-2ex
\end{figure}

The qualitative results of restored PAL images reveal the effectiveness of our ACI framework. Rigorously, we evaluate the framework quantitatively through Modulation Transfer Function (MTF) by $Imatest^\circledR$. As Fig.~\ref{fig:mtf} shows, the black curve is the diffraction limit of selected optical system while the MTF curves of degraded image and restored image are drawn in blue and red, respectively. We conduct the analysis on unfolded images where the MTF is the result of a combination of optical system, ISP and the unfolding process. Without ACI framework, the MTF of unfolded image is far lower than diffraction limit due to the aberration and uneven upsampling, which can also be seen from the blur edge of the test patch at the lower left. However, the MTF of restored image from our proposed method improves the MTF significantly and even exceeds the diffraction limit. The MTF analysis proves that the proposed ACI framework can break the limits of optical design, enabling a PAL system with only 3 spherical lenses to capture clear panoramic images.

\section{Conclusion and Future Work}

In this paper, we explore the trade-off of the high-quality imaging and compact PAL design by proposing the ACI framework. As a flexible high-level framework, our ACI can be deployed to any light-weight PAL system for higher imaging quality in volume-critical scenes. We expect our work paves the way for the subsequent advancement of practical panoramic vision tasks.
 
Specifically, we first obtain degraded PAL unfolded images through a wave-based simulation pipeline. For the target PAL lens, we create the DIVPano dataset and impose various aberration modulation to enhance the robustness of the model. We then propose {\rm PI$^{2}$RNet} to recover degraded images, which utilizes the physical priors of image degradation to assist the recovery process. Both quantitative and qualitative experiments show that our method outperforms current state-of-the-art methods in the PAL image restoration task.

Our framework, with the random distribution of aberrations, is proved to be robust to small vibration in the optical system, \eg~the error of assembly and manufacture. However, the simulation and training process of ACI should be re-conducted for new hardware configuration, \eg~new prototype PAL with different types or numbers of lenses, whose aberrations distribution varies greatly from that of applied system. This is also a limitation of most CI pipelines for aberration correction of optical lens.

In the future, we will focus on panoramic imaging with higher quality in terms of resolution, noise and dynamic range. Further researches on color correction and exposure of imaging sensor are also necessary for impressive imaging results. We are particularly interested in the trade-off between sharpness and noise level of restored PAL images. Moreoever, we will investigate to establish a pre-training model library for improving the generalization ability of our framework to different designs of PAL. In addition, an adaptation of ACI framework to annular PAL images will be explored by shifting the data domain, which is meaningful to some panoramic CV tasks such as panoramic SLAM. 

\section*{Acknowledgments}
The authors would like to thank Dr. Shiqi Chen from Zhejiang University for discussion on imaging simulation.

{\small
\bibliographystyle{IEEEtran}
\normalem
\bibliography{reference}
}

\end{document}